\documentclass[final,authoryear,3p,times]{elsarticle}
\usepackage{multirow,setspace,times,amssymb,amsmath,graphicx,color,rotating,subfigure,url}
\usepackage{lineno}
\usepackage{dcolumn,multirow}
\usepackage{natbib}

\journal{Journal of Economic Dynamics \& Control} 

\begin{document}

\begin{frontmatter}

\title{An empirical behavioural order-driven model with price limit rules}
\author[SB,RCE]{Gao-Feng Gu}
\author[CME,CCSCA]{Xiong Xiong}
\author[SB,RCE]{Hai-Chuan Xu}
\author[CME,CCSCA]{Wei Zhang}
\author[CME,CCSCA]{Yong-Jie Zhang}
\author[SZSE]{Wei Chen}
\author[SB,RCE,SS]{Wei-Xing Zhou\corref{cor}}
\cortext[cor]{Corresponding author. Address: 130 Meilong Road, P.O. Box 114, School of Business, East China University of Science and Technology, Shanghai 200237, China, Phone: +86 21 64253634.}
\ead{wxzhou@ecust.edu.cn} %

\address[SB]{Department of Finance, East China University of Science and Technology, Shanghai 200237, China}
\address[RCE]{Research Center for Econophysics, East China University of Science and Technology, Shanghai 200237, China}
\address[CME]{College of Management and Economics, Tianjin University, Tianjin 300072, China}
\address[CCSCA]{China Center for Social Computing and Analytics, Tianjin University, Tianjin 300072, China}
\address[SZSE]{Shenzhen Stock Exchange, 5045 Shennan East Road, Shenzhen 518010, China}
\address[SS]{Department of Mathematics, East China University of Science and Technology, Shanghai 200237, China}

\begin{abstract}
  We develop an empirical behavioural order-driven (EBOD) model, which consists of an order placement process and an order cancellation process. Price limit rules are introduced in the definition of relative price. The order placement process is determined by several empirical regularities: the long memory in order directions, the long memory in relative prices, the asymmetric distribution of relative prices, and the nonlinear dependence of the average order size and its standard deviation on the relative price. Order cancellation follows a Poisson process with the arrival rate determined from real data and the cancelled order is determined according to the empirical distributions of relative price level and relative position at the same price level. All these ingredients of the model are derived based on the empirical microscopic regularities in the order flows of stocks on the Shenzhen Stock Exchange. The model is able to produce the main stylized facts in real markets. Computational experiments uncover that asymmetric setting of price limits will cause the stock price diverging exponentially when the up price limit is higher than the down price limit and vanishing vice versus. We also find that asymmetric price limits have influences on stylized facts. Our EBOD model provides a suitable computational experiment platform for academics, market participants and policy makers.
\end{abstract}


\end{frontmatter}

\section{Introduction}
\label{se:introduction}

The emerging Chinese stock market adopts the price limit mechanism since 2 January 1997 to refrain speculative behaviours and stabilize the market. The mechanism sets symmetric price limits in which the up limit $\phi_+=10\%$ and the down limit $\phi_-=-10\%$ for common stocks and $\phi_+=5\%$ and $\phi_-=-5\%$ for specially treated (ST and *ST) stocks. On 27 July 2012 (Friday), after the closing of the Chinese stock market, the Shanghai Stock Exchange (SHSE) released the ``Draft Guideline on the Trading of Stocks Bearing Risk Warnings''. A key term (Article VII) suggested that, for stocks bearing risk warnings, the maximum percent of price increase is $\phi_+=1\%$, while the maximum percent of price decrease is $\phi_-=-5\%$. Affected by this event, on 30 July 2012, 106 out of the 110 ST and *ST shares were sealed at the down price limit. On that day, the SHSZ Composite Index, the Shenzhen Stock Exchange (SZSE) Composite Index and the Chinese Stock Market 500 Index (CSI 500) dropped respectively $-0.89\%$, $-1.65\%$ and $-2.01\%$.

This Draft, especially Article VII, incurred a lot of opposition. Yu Chen, the general manager of Beijing Shennong Capital Management, released an open letter to the SHSE, objecting the asymmetric price limit setting on risk-warning stocks. He argued that the Exchange should not enable unfair trading rules since the provision of asymmetric price limits will significantly increase the cost of buying and holding. He urged that Article VII should be removed from the Draft. The final Guideline did replace the asymmetric price limits with symmetric limits. The opinions of economists and market participants are certainly correct theoretically. However, is there any technical evidence supporting these ``theoretical arguments''? In this work, we design an empirical behavioural order-driven model. Armed with this new model, computational experiments show that the asymmetric setting with larger down limit ($\phi_+<|\phi_-|$) leads to vanishing prices. Hence, under the Draft Guideline, any stock cannot escape the fate of delisting once it is labelled risk warning.

The main motivation of setting price limits is to curb speculative trading and it is expected to have a cooling-off effect to reduce the volatility of securities \citep{Ma-Rao-Sears-1989-JFSR}. Nevertheless, there are also theoretical analyses predicting the presence of a magnet effect, which refers to the phenomenon that the price limit acts as a magnet to attract more orders towards the same-side price limit when the price approaches either limit \citep{Subrahmanyam-1994-JF}. The study of price limits can be traced back to the 1980's \citep{Telser-1981-JFutM,Brennan-1986-JFE}. However, the presence of a magnet or cooling-off effect and the effectiveness of the price limit rules are controversial. While \cite{Du-Liu-Rhee-2005-WP}, \cite{Hsieh-Kim-Yang-2009-JEF}, and \cite{Wong-Liu-Zeng-2009-CER} find a magnet effect on both price limits, \cite{Arak-Cook-1997-JFSR} and \cite{Wan-Xie-Gu-Jiang-Chen-Xiong-Zhang-Zhou-2015-PLoS1} find a cooling-off effect. In contrast, \cite{Cho-Russell-Tiao-Tsay-2003-JEF}, \cite{Chen-Yang-Fan-2008-cnJMSC}, and \cite{Li-Geng-2012-cnRIS} report a magnet effect towards the up-limit and an insignificant or weak magnet effect towards the down-limit, and \cite{Zhang-Zhu-2014-cnJCQUT} present a magnet effect of the up-limit and a cooling-off effect of the down-limit.

There are studies that design agent-based models and perform computational experiments to investigate the effects of price limits. \cite{Yeh-Yang-2013-JEIC} find no evidence of volatility spillover but phenomena of delayed price discovery and trading interference, whose significance depends on the level of the price limits. \cite{Zhang-Ping-Zhu-Li-Xiong-2016-PLoS} demonstrate that both price limits can cause a volatility spillover effect and a trading interference effect. Moreover, \cite{Yeh-Yang-2010-JEDC} and \cite{Xiong-Ding-Yang-Zhang-2015-PLoS} argue that a proper level of price limits is helpful to stabilize the markets. Inspired by the asymmetric effect of symmetric price limit mechanism in the Chinese stock market, \cite{Li-Geng-2012-cnRIS} perform Monte Carlo simulations of the VF-EGARCH-M model and conclude that there is an optimal design of asymmetric price limit mechanism with $\phi_+>|\phi_-|$. However, our computational experiments show that asymmetric price limit rules do hurt the market.

The Chinese stock market is an order-driven market, which adopts the continuous double auction mechanism and symmetric price limits mechanism \citep{Gu-Chen-Zhou-2007-EPJB,Jiang-Chen-Zhou-2008-PA,Mu-Chen-Kertesz-Zhou-2009-EPJB,Wu-Wang-Li-2017-PA}. The model proposed in this work was inspired by the seminal Mike-Farmer model of \cite{Mike-Farmer-2008-JEDC} and an updated version of \cite{Gu-Zhou-2009-EPL}'s model. This family of order-driven models mimics the order placement process and order cancellation process entering the trading host. The order placement process contains three components: order direction, relative order price and order size. The microscopic rules of these models are obtained by empirical regularities of these processes. Compared with existing models \citep{Mike-Farmer-2008-JEDC,Gu-Zhou-2009-EPL}, the new model makes two innovations. First, it adopts a new definition of relative order price by integrating price limits $\phi_+$ and $\phi_-$ as model parameters. This enables us to investigate the mechanisms of phenomena caused by price limit that we will review briefly below. Second, it considers order sizes. Hence, the model is more realistic and significantly enhances its ability for the study of market microstructure theories such as immediate price impacts \citep{Lillo-Farmer-Mantegna-2003-Nature,Lim-Coggins-2005-QF,Zhou-2012-NJP,Zhou-2012-QF,Xu-Jiang-Zhou-2017-IJMPB,Pham-Duong-Lajbcygier-2017-JFc}. The models of \cite{Mike-Farmer-2008-JEDC} and \cite{Gu-Zhou-2009-EPL} have been applied to understand the mechanisms underlying stylized facts \citep{Gu-Zhou-2009-EPJB,Meng-Ren-Gu-Xiong-Zhang-Zhou-Zhang-2012-EPL,Zhou-Gu-Jiang-Xiong-Chen-Zhang-Zhou-2017-CE}.

The study of order-driven models has a long history, which can be referred to \cite{Stigler-1964-JB} more than 50 years ago. Researchers have construct different order-driven models based on different groups of micro driving rules, attempting to simulate the dynamics of the limit-order book \citep{Maslov-2000-PA,Farmer-Patelli-Zovko-2005-PNAS,Mike-Farmer-2008-JEDC,Gu-Zhou-2009-EPL,Tseng-Lin-Lin-Wang-Li-2010-PA}. In these models, orders are not specific to certain traders and the traders have zero intelligence. Other behavioural models include percolation models \citep{Stauffer-1998-AP,Cont-Bouchaud-2000-MeD,Eguiluz-Zimmermann-2000-PRL}, Ising models
\citep{Foellmer-1974-JMathE,Chowdhury-Stauffer-1999-EPJB,Iori-1999-IJMPC,Kaizoji-2000-PA,Bornholdt-2001-IJMPC,Zhou-Sornette-2007-EPJB}, minority games
\citep{Arthur-1994-AER,Challet-Zhang-1997-PA,Challet-Marsili-Zhang-2000-PA,Challet-Marsili-Zhang-2001a-PA,Challet-Marsili-Zhang-2001b-PA,Challet-Marsili-Zhang-2005}, and so on. Another important family is heterogenous agent models, in which the agents are combinations of informed traders, fundamentalists, technical traders, smart traders, noise traders, and so on \citep{Brock-Hommes-1997-Em,Brock-Hommes-1998-JEDC,Lux-Marchesi-1999-Nature,Chiarella-Iori-2002-QF,Chiarella-He-Wang-2006-CSF,Barunik-Vacha-Vosvrda-2009-JEIC,Xu-Zhang-Xiong-Zhou-2014-AAA,Xu-Zhang-Xiong-Zhou-2014-MPE}. Under this framework, one can study the effects of diverse factors, such as noise \citep{Chiarella-He-Zhang-2011-JEDC}, technical trading rules \citep{Chiarella-Iori-Perello-2009-JEDC,He-Li-2015-JBF}, and investor sentiment \citep{Chiarella-He-Shi-Wei-2017-QF}. Excellent reviews have been provided by \cite{Chakraborti-Toke-Patriarca-Abergel-2011b-QF} and \cite{Sornette-2014-RPP}. Nevertheless, the effects of asymmetric price limits have not been studied with computational experiments within these models.

The rest of paper is organized as follows. Section~\ref{se:dateset} describes briefly the database we adopt. Section~\ref{sec:modeldetail} constructs the behavioural order-driven model based on empirical regularities of order flows. We perform computational experiments in Section~\ref{sec:meodelapp} to study the effects of asymmetric price limit rules on the evolution of stock price and several stylized facts. Section~\ref{sec:conclusion} summarizes the results.

\section{Date sets}
\label{se:dateset}

We use the ultra-high-frequency order flow data of 32 A-share stocks and 11 B-share stocks traded on the Shenzhen Stock Exchange (SZSE) in 2003 to build the order-driven model. The 32 A-shares were constituents of the Shenzhen Stock Exchange Component Index and the 11 B-share stocks paired with 11 A-share stocks in our sample.

Each entry of the records contains the details of order placement and order cancellation, including the order placement/cancellation time, order price, order size and order identifier which identifies whether the order is a buy order, a sell order, or a cancellation. The time stamp of the database is accurate to 0.01s. The data allow us to reconstruct the limit order books (LOBs) and reproduce the price formation process.

\section{Order-driven model description}
\label{sec:modeldetail}

The placed orders can be regarded as contracts for investors willing to buy or sell certain stock shares at certain price. Once an order is placed, it will organized in a queue in the LOB. It is clear that the LOB has two opposite sides, that is, buy LOB and sell LOB. Buy orders in the buy LOB is arranged by decreasing the order price and the highest price of the order at the top is called best bid. In the sell LOB, sell orders are arranged by increasing the order price and the lowest price at the top is called best ask or best offer. If orders have the same order price, they will be arranged based on the placement time. Orders arriving earlier have the priority to be executed.

Order cancellation also plays an important role in the price formation of the stock market. If orders are not fully filled, they are usually cancelled from the LOB based on investor decision. When limit orders at the best price cancelled completely, mid-price and spread will change as well. If cancellation occurs inside the LOB, it also affects the shape of LOB and has potential effects on price formation.

The prices are formed due to the order placement and cancellation processes. Hence, the order-driven model contains these two independent processes. In the order placement process, three ingredients of an order are considered: order direction, order price and order size.

\subsection{Order direction}

The first ingredient of order placement is the order direction $s$. Assuming $s=+1$ for buy orders and $s=-1$ for sell orders, we can construct the order direction series for each stock. To measure the memory effect of the order direction series for all the 43 stocks, we apply the Detrending Moving Average Analysis (DMA) \citep{Carbone-2009-IEEE,Gu-Zhou-2010-PRE}, which is among the best estimators of Hurst exponent \citep{Jiang-Zhou-2011-PRE,Shao-Gu-Jiang-Zhou-Sornette-2012-SR}. The detrending fluctuation function $F(\ell)$ can be computed and is expected to be power-law related to the scale size $\ell$, which reads
\begin{equation}
 F(\ell) \sim \ell^H~,
 \label{Eq:dma-fs}
\end{equation}
where $H$ is known as the DMA scaling exponent or roughly the Hurst exponent. The time series is persistent if $H>0.5$, uncorrelated if $H=0.5$, and antipersistent if $H<0.5$. A persistent time series has long memory. We use $H_s$ with subscript $s$ for the Hurst exponents of order directions.

\begin{figure}[htb]
  \centering
  \includegraphics[width=7.5cm,height=6cm]{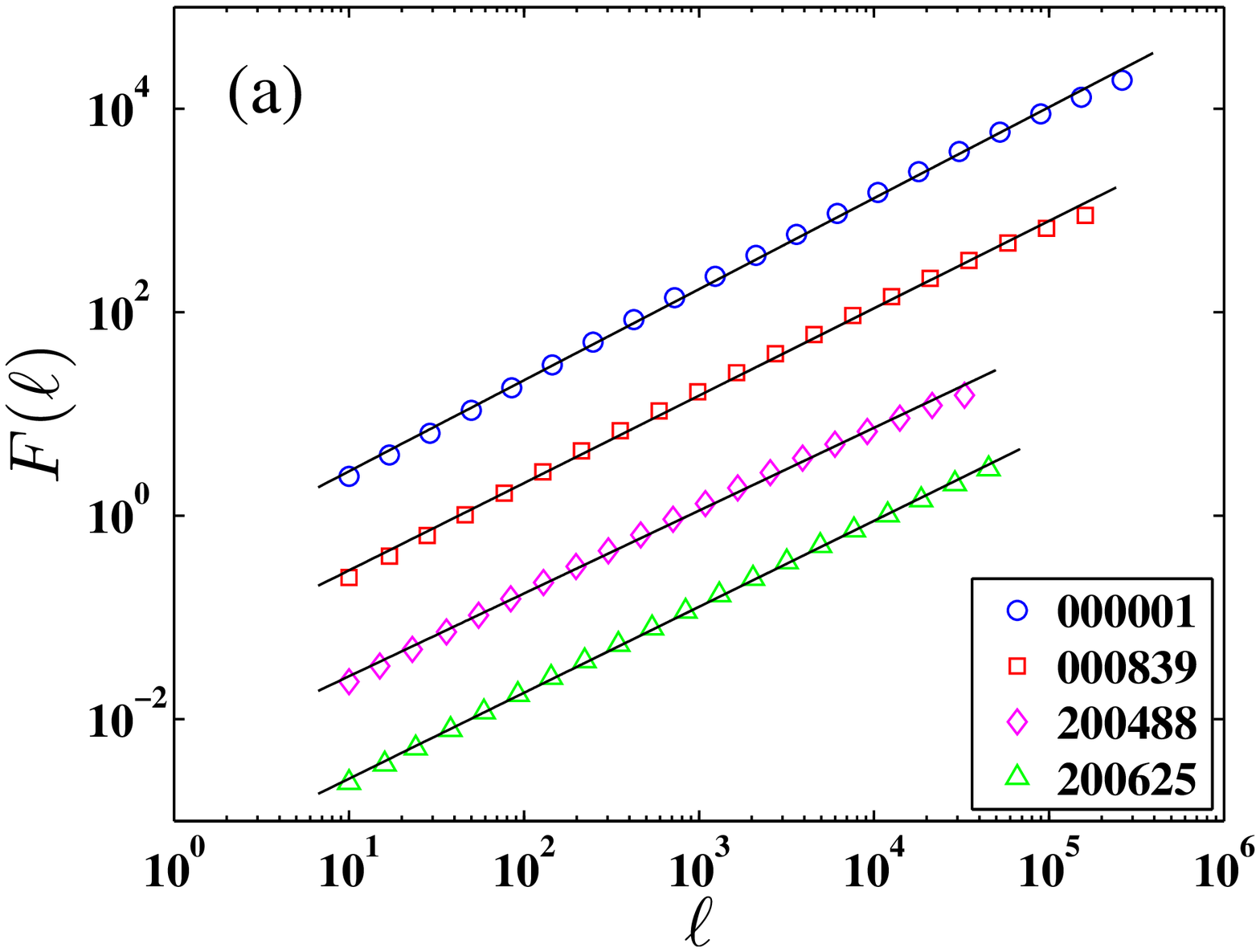}
  \includegraphics[width=7.5cm,height=6cm]{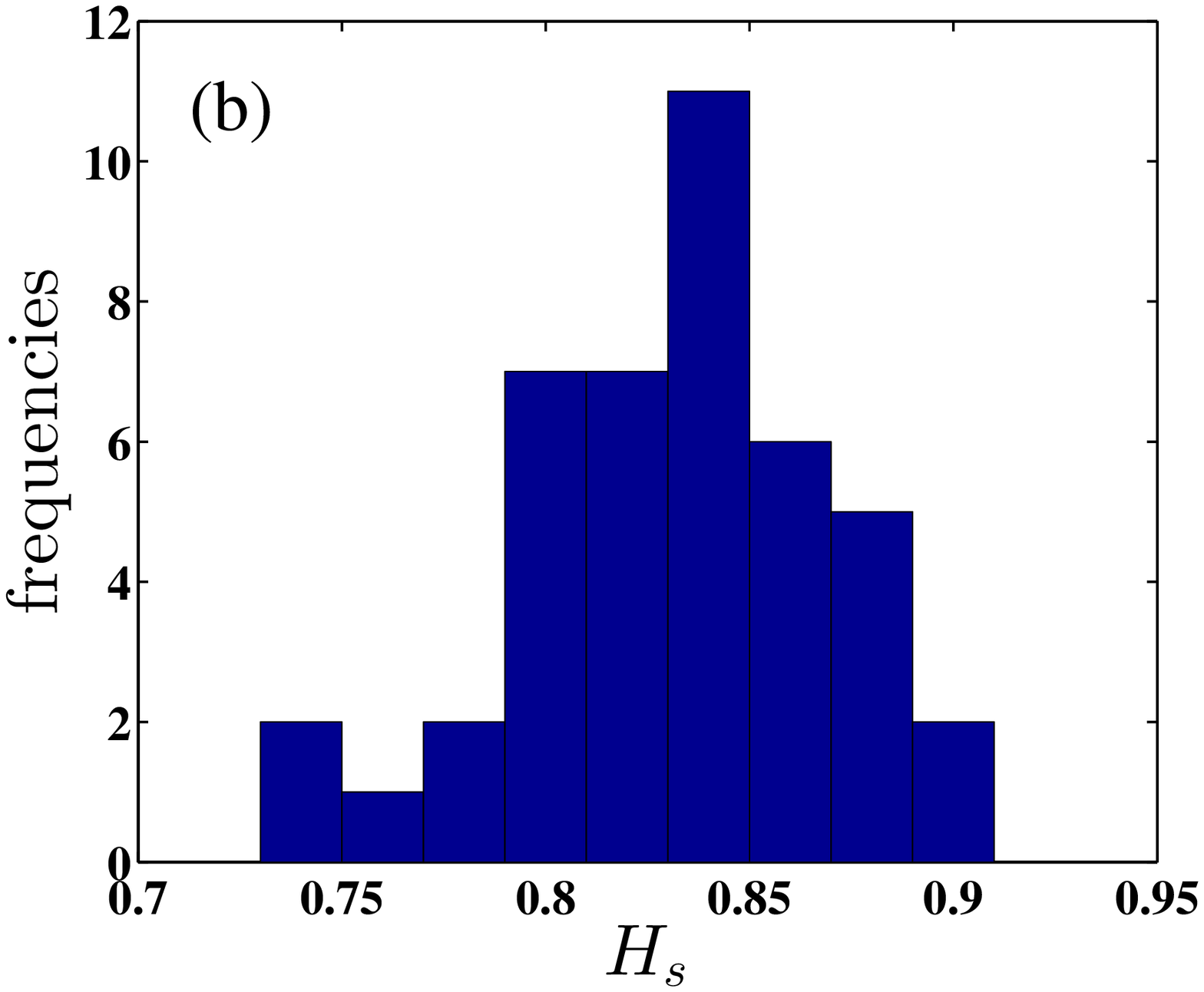}
  \caption{(colour online) Long memory in order directions. (a) Plots of the fluctuation functions $F(\ell)$ of order directions for four stocks 000001, 000839, 200488 and 200625. The solid lines are the least squares fits to the data. The scaling curves for 000839, 200488 and 200625 have been shifted vertically for clarity. (b) Histograms of Hurst exponents $H_s$ of order directions for all the 43 stocks.}
  \label{Fig:EBOD:Rules:OrderDirection}
\end{figure}

The fluctuation functions $F(\ell)$ with respect to the scale $\ell$ of four stocks (000001 and 000839 are A shares and 200488 and 200625 are B shares) are illustrated in Fig.~\ref{Fig:EBOD:Rules:OrderDirection}(a). Excellent power law scaling relations are observed in the scaling ranges over 4 to 5 orders of magnitude. The Hurst exponents of the four stocks are estimated to be $H_s=0.895 \pm 0.008$ for stock 000001, $H_s=0.859 \pm 0.009$ for stock 000839, $H_s=0.814 \pm 0.009$ for stock 200488, $H_s=0.845 \pm 0.005$ for stock 200625. The Hurst exponents of all the 43 stocks are presented in Table~\ref{Tb:EBOD:Hurst:OrderDirection}. We also show the histogram of Hurst exponents of all the stocks in Fig.~\ref{Fig:EBOD:Rules:OrderDirection}(b). It is clear that all the Hurst exponents are all larger than 0.5 and we document that the order direction series significantly has long memory. According to Table~\ref{Tb:EBOD:Hurst:OrderDirection}, the mean Hurst exponent for all the stocks is $\langle{H_s}\rangle=0.841 \pm 0.036$. Moreover, the mean Hurst exponent of A shares is $\langle{H_{s,A}}\rangle=0.854\pm0.027$, which is little larger than that of B shares $\langle{H_{s,B}}\rangle=0.806\pm0.036$. It indicates that stronger persistence exists in the A shares in the year of 2003 in the Chinese stock markets. The results are consistent with the finding of \cite{Lillo-Farmer-2004-SNDE} for London Stock Exchange (LSE) stocks. However, we find that the SZSE stocks have stronger persistence in order directions than the LSE stocks, which reflects the fact that there are stronger imitative and herding behaviours in the Chinese stock market.

\begin{table}[htp]
  \centering
  \caption{Hurst exponents $H_s$ of order directions for the 32 A-share stocks and 11 B-share stocks estimated using the detrending moving average (DMA) method. The mean Hurst exponent is $\langle{H_s}\rangle=0.841 \pm 0.036$ for all stocks, $\langle{H_{s,A}}\rangle=0.854 \pm 0.027$ for A shares, and $\langle{H_{s,B}}\rangle=0.806\pm 0.036$ for B-shares.}
  \medskip
  \label{Tb:EBOD:Hurst:OrderDirection}
  \centering
  \begin{tabular}{cccc|cc|cc}
  \hline \hline
   Stock & $H_{s,A}$ & Stock & $H_{s,B}$  & Stock & $H_{s,A}$  & Stock & $H_{s,A}$  \\
  \hline
    000002 & 0.904 $\pm$ 0.005 & 200002 & 0.844 $\pm$ 0.001 & 000001 & 0.895 $\pm$ 0.008 & 000778 & 0.842 $\pm$ 0.005 \\
    000012 & 0.834 $\pm$ 0.007 & 200012 & 0.812 $\pm$ 0.003 & 000009 & 0.866 $\pm$ 0.010 & 000800 & 0.871 $\pm$ 0.005 \\
    000016 & 0.841 $\pm$ 0.005 & 200016 & 0.759 $\pm$ 0.005 & 000021 & 0.824 $\pm$ 0.010 & 000825 & 0.906 $\pm$ 0.003 \\
    000024 & 0.832 $\pm$ 0.005 & 200024 & 0.751 $\pm$ 0.009 & 000027 & 0.891 $\pm$ 0.006 & 000839 & 0.859 $\pm$ 0.009 \\
    000429 & 0.827 $\pm$ 0.003 & 200429 & 0.816 $\pm$ 0.004 & 000063 & 0.856 $\pm$ 0.008 & 000858 & 0.885 $\pm$ 0.003 \\
    000488 & 0.814 $\pm$ 0.004 & 200488 & 0.814 $\pm$ 0.009 & 000066 & 0.829 $\pm$ 0.009 & 000898 & 0.884 $\pm$ 0.007 \\
    000539 & 0.852 $\pm$ 0.004 & 200539 & 0.870 $\pm$ 0.002 & 000088 & 0.844 $\pm$ 0.005 & 000917 & 0.815 $\pm$ 0.006 \\
    000541 & 0.862 $\pm$ 0.002 & 200541 & 0.774 $\pm$ 0.003 & 000089 & 0.864 $\pm$ 0.003 & 000932 & 0.876 $\pm$ 0.005 \\
    000550 & 0.834 $\pm$ 0.007 & 200550 & 0.781 $\pm$ 0.009 & 000406 & 0.830 $\pm$ 0.008 & 000956 & 0.848 $\pm$ 0.007 \\
    000581 & 0.892 $\pm$ 0.002 & 200581 & 0.799 $\pm$ 0.006 & 000709 & 0.860 $\pm$ 0.006 & 000983 & 0.853 $\pm$ 0.004 \\
    000625 & 0.806 $\pm$ 0.009 & 200625 & 0.845 $\pm$ 0.005 & 000720 & 0.819 $\pm$ 0.002 &        &                   \\
  \hline\hline
 \end{tabular}
\end{table}

\subsection{Relative price}

We define the relative price $x$ based on the price limit rules in the Chinese stock market, which is different from the definition of \cite{Mike-Farmer-2008-JEDC} for LSE stocks. We require that the relative price $x$ varies in the range $[-1,1]$ and stands for the order aggressiveness. When a buy order is placed on the down limit or a sell order is placed on the up limit, we have $x=-1$. These orders are the least aggressive. When a buy order is placed on the up limit or a sell order is placed on the down limit, we have $x=1$. These orders are the most aggressive. When an order is place on the opposite best price, i.e. $p_t=p_a(t-1)$ for buy orders and $p_t=p_b(t-1)$ for sell orders, we have $x=0$. Following these considerations, we define the relative price as follows:
\begin{equation}
  x_t=\left\{
  \begin{array}{ll}
    \left[ p_b(t-1)-p_t \right] / \left[ p_b(t-1)-p_{\rm{min}}(T) \right]&~~~{\mathrm{for~sell~market~orders}} \\
    \left[ p_b(t-1)-p_t \right] / \left[ p_{\rm{max}}(T)-p_b(t-1) \right]&~~~{\mathrm{for~sell~limit~orders}}\\
    \left[ p_t-p_a(t-1) \right] / \left[ p_{\rm{max}}(T)-p_a(t-1) \right]&~~~{\mathrm{for~buy~market~orders}} \\
    \left[ p_t-p_a(t-1) \right] / \left[ p_a(t-1)-p_{\rm{min}}(T) \right]&~~~{\mathrm{for~buy~limit~orders}}
  \end{array}
  \right.,
  \label{Eq:EBOD:x}
\end{equation}
where $p_t$ is the price of order placed at event time $t$, $p_a(t-1)$ and $p_b(t-1)$ are the best ask and best bid at event time $t-1$, and $p_{\rm{max}}(T)$ and $p_{\rm{min}}(T)$ are the maximum and minimum valid prices on a trading day $T$. Larger $x$ values impliy that traders are more eager to make a transaction immediately and the orders are thus more aggressive. Effective market orders ($x \geq 0$) result in an immediate transaction, while effective limit orders ($x < 0$) are stored in the limit order book waiting to be executed by future effective market orders on the opposite side.

If the price is at the up limit when an order arrives at $t$, we have
\begin{equation}
  p_a(t-1)=p_b(t-1)=p_{\rm{max}}(T).
  \label{Eq:EBOD:pa:pb:uplimit}
\end{equation}
Under such scenarios, the relative prices of sell market orders and buy limit orders are well defined by Eq.~(\ref{Eq:EBOD:x})
\begin{equation}
  x_t=\left\{
  \begin{array}{ll}
    \left[ p_b(t-1)-p_t \right] / \left[ p_b(t-1)-p_{\rm{min}}(T) \right]&~~~{\mathrm{for~sell~market~orders}} \\
    \left[ p_t-p_a(t-1) \right] / \left[ p_a(t-1)-p_{\rm{min}}(T) \right]&~~~{\mathrm{for~buy~limit~orders}}
  \end{array}
  \right.,
  \label{Eq:EBOD:x:uplimit:1}
\end{equation}
while the relative prices of buy market orders and sell limit orders cannot be defined by Eq.~(\ref{Eq:EBOD:x}). For buy market orders, one requires that $p_t>p_b(t)=p_{\rm{max}}(T)$, which is however not permitted. For sell limit orders, one requires $p_t>p_a(t)=p_{\rm{max}}(T)$, which is again not permitted. In other words, when the price is at the up limit, both buy market orders and sell limit orders do not exist. Hence, we simply pose that
\begin{equation}
  p_t= p_{\rm{max}}(T)
  \label{Eq:EBOD:x:uplimit:2}
\end{equation}
no matter what the value of $x>0$ is. The situation for down price limit is similar. If the price is at the down limit when an order arrives at $t$, we have
\begin{equation}
  p_a(t-1)=p_b(t-1)=p_{\rm{min}}(T).
  \label{Eq:EBOD:pa:pb:downlimit}
\end{equation}
Under such scenarios, the relative prices of buy market orders and sell limit orders are well defined by Eq.~(\ref{Eq:EBOD:x})
\begin{equation}
  x_t=\left\{
  \begin{array}{ll}
    \left[ p_b(t-1)-p_t \right] / \left[ p_{\rm{max}}(T)-p_b(t-1) \right]&~~~{\mathrm{for~sell~limit~orders}}\\
    \left[ p_t-p_a(t-1) \right] / \left[ p_{\rm{max}}(T)-p_a(t-1) \right]&~~~{\mathrm{for~buy~market~orders}}
  \end{array}
  \right.,
  \label{Eq:EBOD:x:downlimit:1}
\end{equation}
while the relative prices of sell market orders and buy limit orders cannot be defined by Eq.~(\ref{Eq:EBOD:x}). For sell market orders, one requires that $p_t<p_a(t)=p_{\rm{min}}(T)$, which is however not permitted. For buy limit orders, one requires $p_t<p_a(t)=p_{\rm{min}}(T)$, which is also not permitted. In other words, when the price is at the down limit, both sell market orders and buy limit orders do not exist. Therefore, we simply pose that
\begin{equation}
  p_t= p_{\rm{min}}(T),
  \label{Eq:EBOD:x:downlimit:2}
\end{equation}
regardless of what the value of $x<0$ is.

\begin{figure}[htb]
  \centering
  \includegraphics[width=7.5cm,height=6cm]{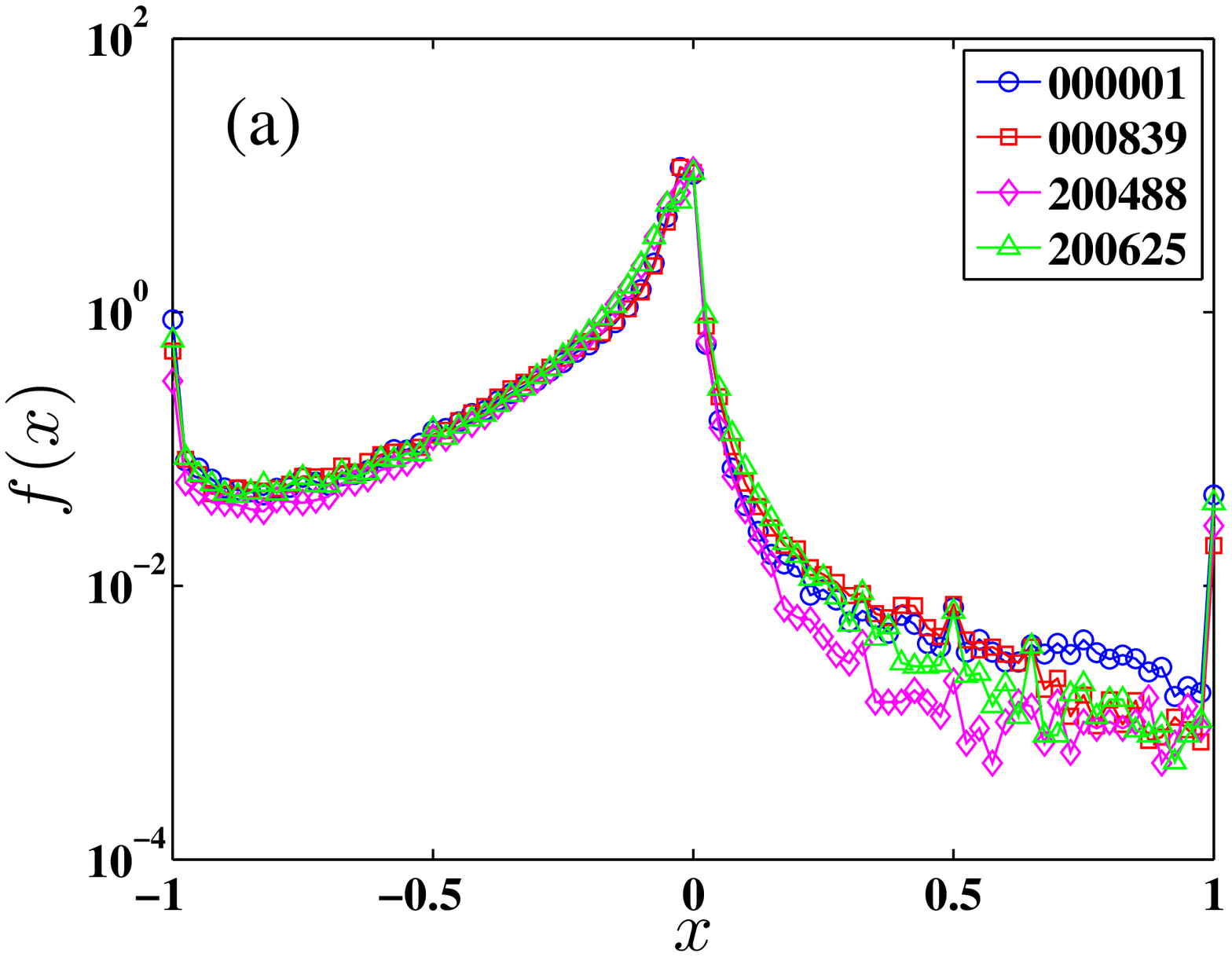}
  \includegraphics[width=7.5cm,height=6cm]{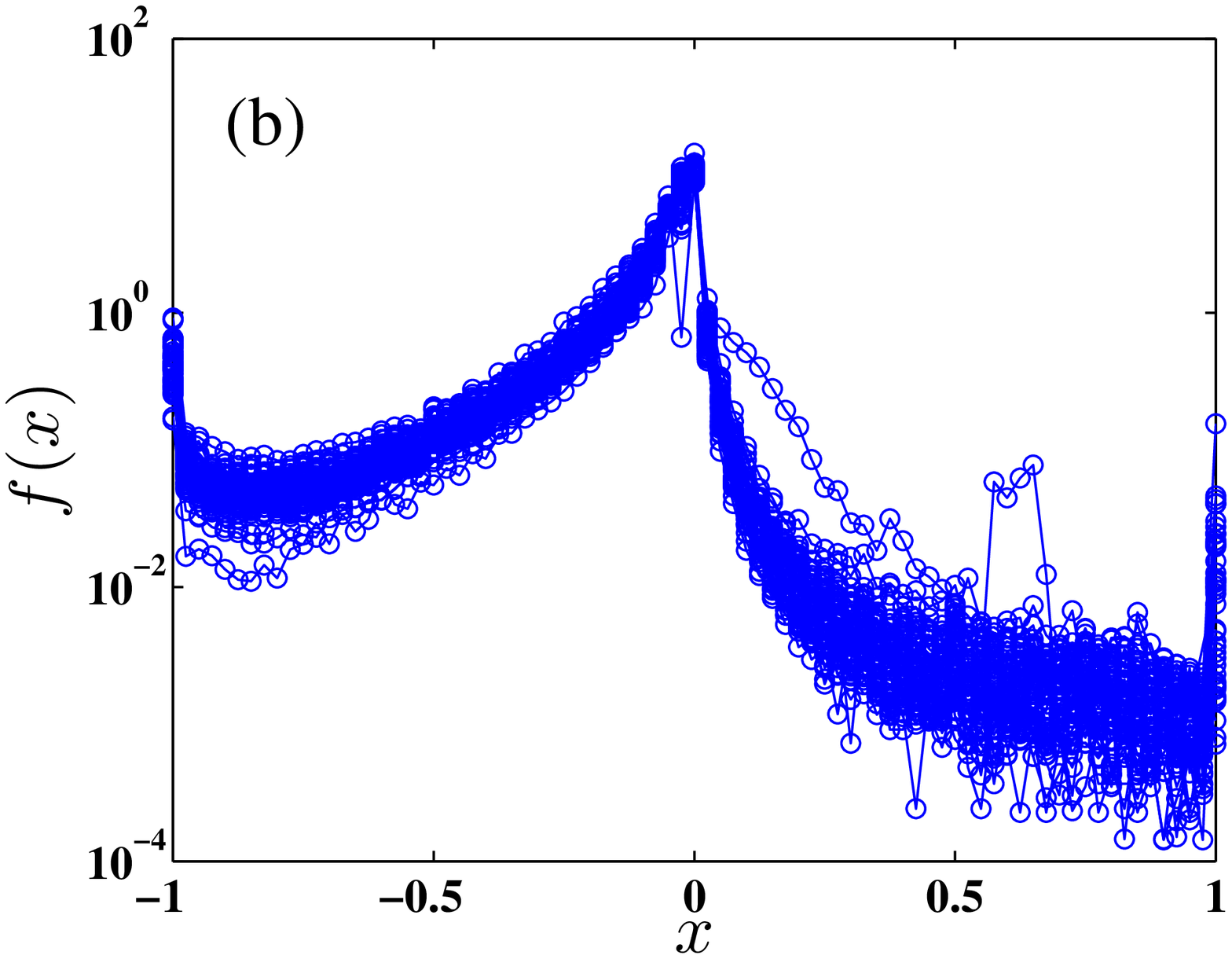}
  \includegraphics[width=7.5cm,height=6cm]{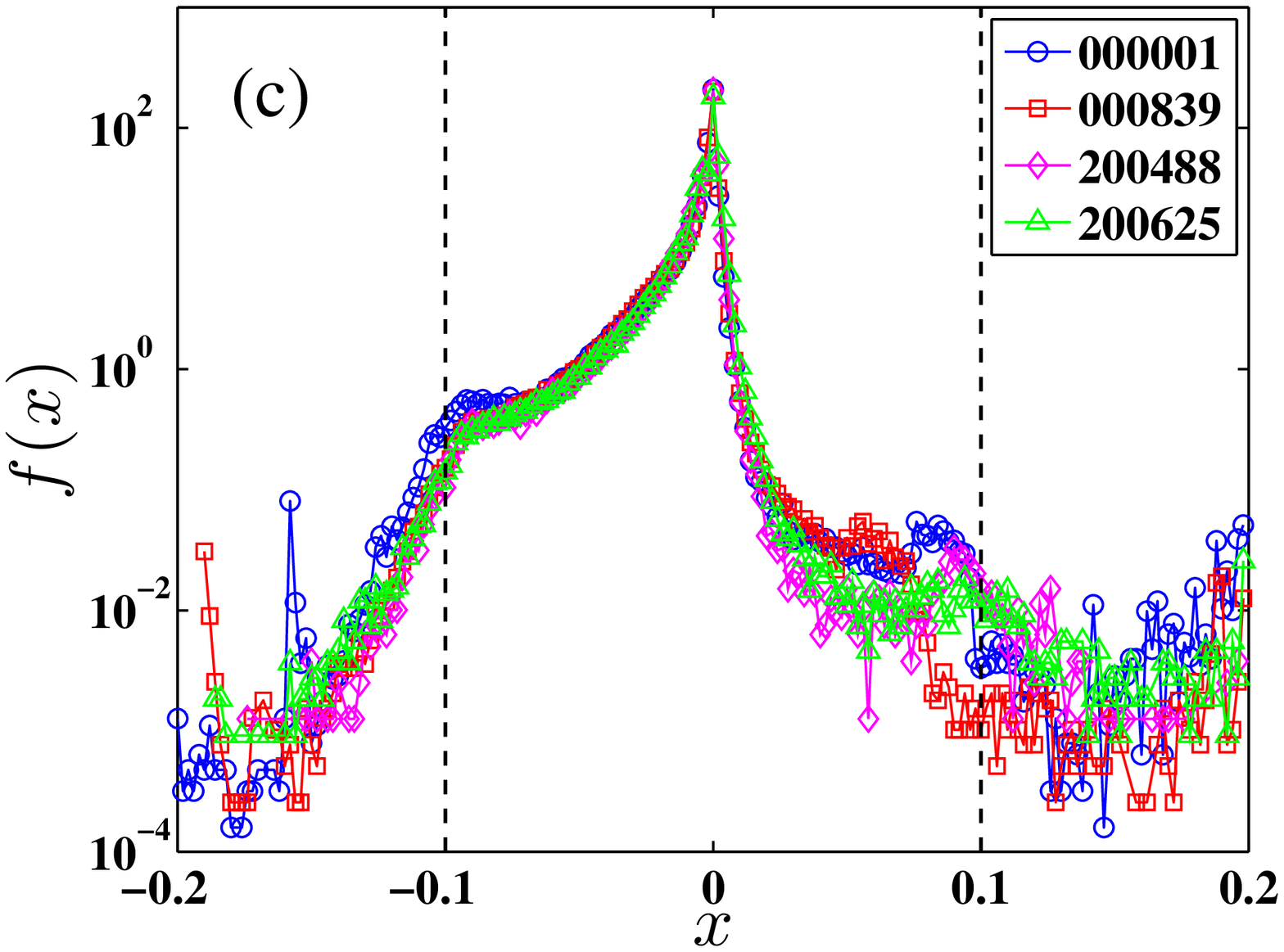}
  \includegraphics[width=7.5cm,height=6cm]{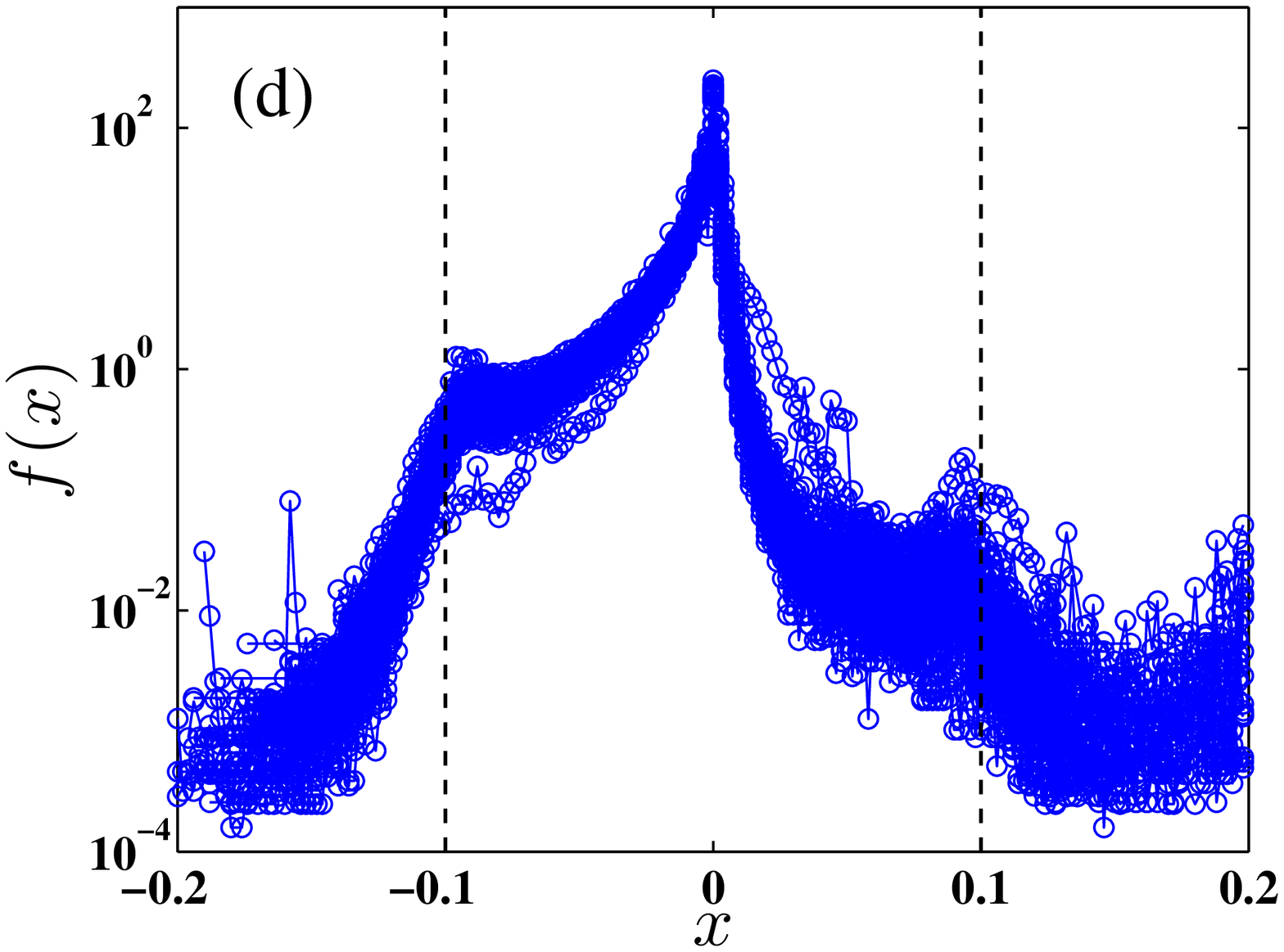}
  \caption{(colour online) Empirical probability density function $f(x)$ of relative prices. (a,c) Two A-share stocks 000001 and 000839 and two B-share stocks 200488 and 200625. (b,d) All the 43 stocks. The relative prices are defined according to price limits in plots (a) and (b) and according to \cite{Mike-Farmer-2008-JEDC} in plots (c) and (d).}
  \label{Fig:EBOD:Price:PDF}
\end{figure}

Fig.~\ref{Fig:EBOD:Price:PDF}(a) presents the empirical probability density functions (PDFs) $f(x)$ of relative prices aggregating both buy and sell orders for four representative stocks in the sample. We find that the PDF curves of four stocks almost collapse together, especially in the range $x<0$. Moreover, other stocks have similar probability distributions as presented in Fig.~\ref{Fig:EBOD:Price:PDF}(b), except for the stock 000720 which has obviously higher probabilities in the range $0 < x < 0.5$. The $f(x)$ functions reach their maximums around the point $x=0$, which means that many traders trend to place orders at the opposite best price in order to balance the relation between the transaction cost and transaction opportunity. The distributions are asymmetric (the skewness equal to $-2.69$ for stock 000001), which implies that more orders are placed in the limit order book. According to the order flow data of stock 000001, only 28.28\% of the placed orders are efficient market orders with $x \geq 0$. This observation is natural to maintain nonempty LOBs. We also find that the values at $x=\pm1$ are significant jumps, which shows that quite a few traders place extreme orders at the down or up limit.
Fig.~\ref{Fig:EBOD:Price:PDF}(c) and Fig.~\ref{Fig:EBOD:Price:PDF}(d) show the empirical probability density functions $f(x)$ of relative prices without price limits defined in \cite{Mike-Farmer-2008-JEDC} for comparison. Under this definition, the relative price $x$ varies in $[-0.2,0.2]$. The most remarkable feature is that there are ``crossovers'' at $x=\pm10\%$, which reflects the effect of price limits. In addition, the jumps at $x=\pm20\%$ are less significant.

\begin{figure}[htb]
  \centering
  \includegraphics[width=7.5cm,height=6cm]{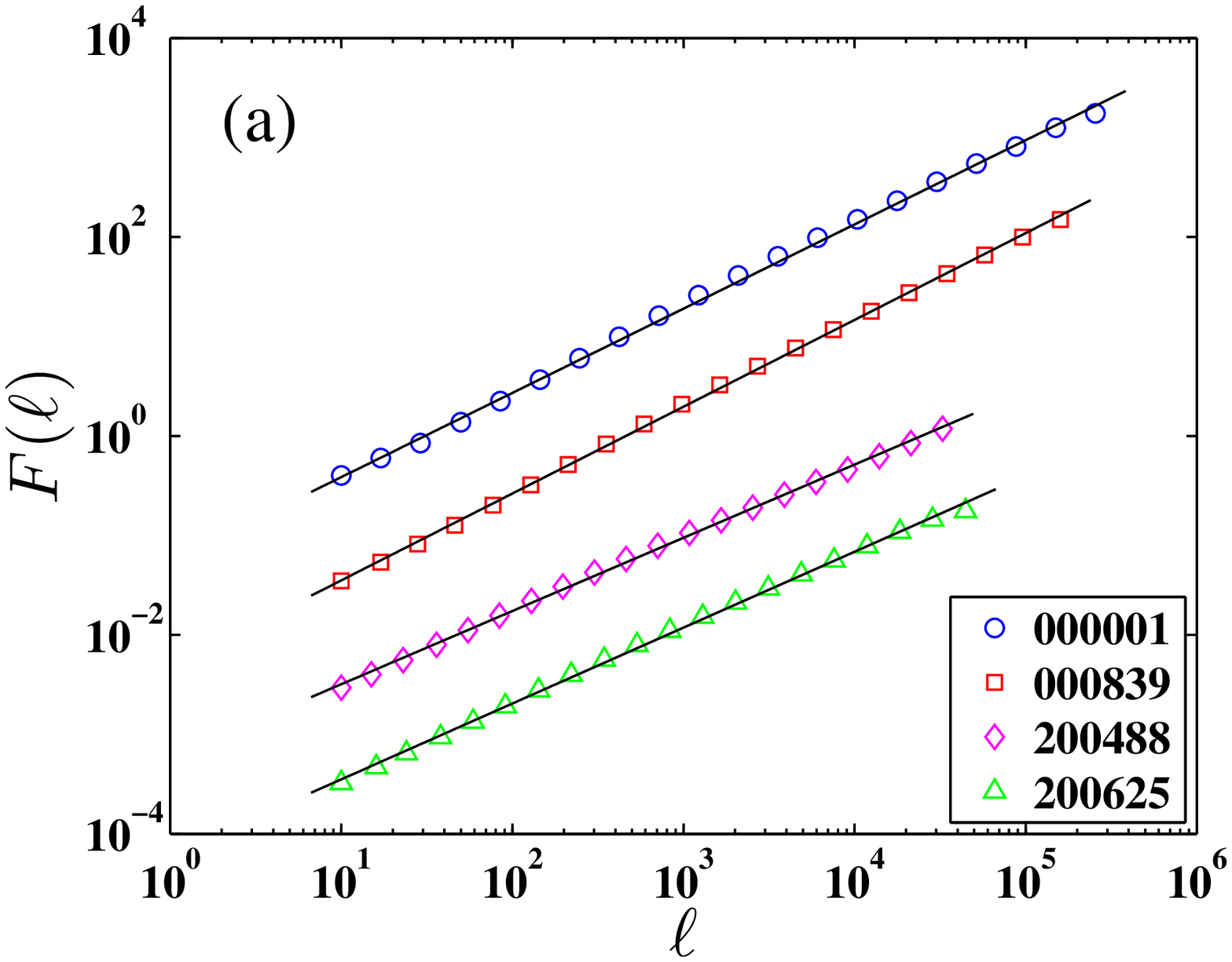}
  \includegraphics[width=7.5cm,height=6cm]{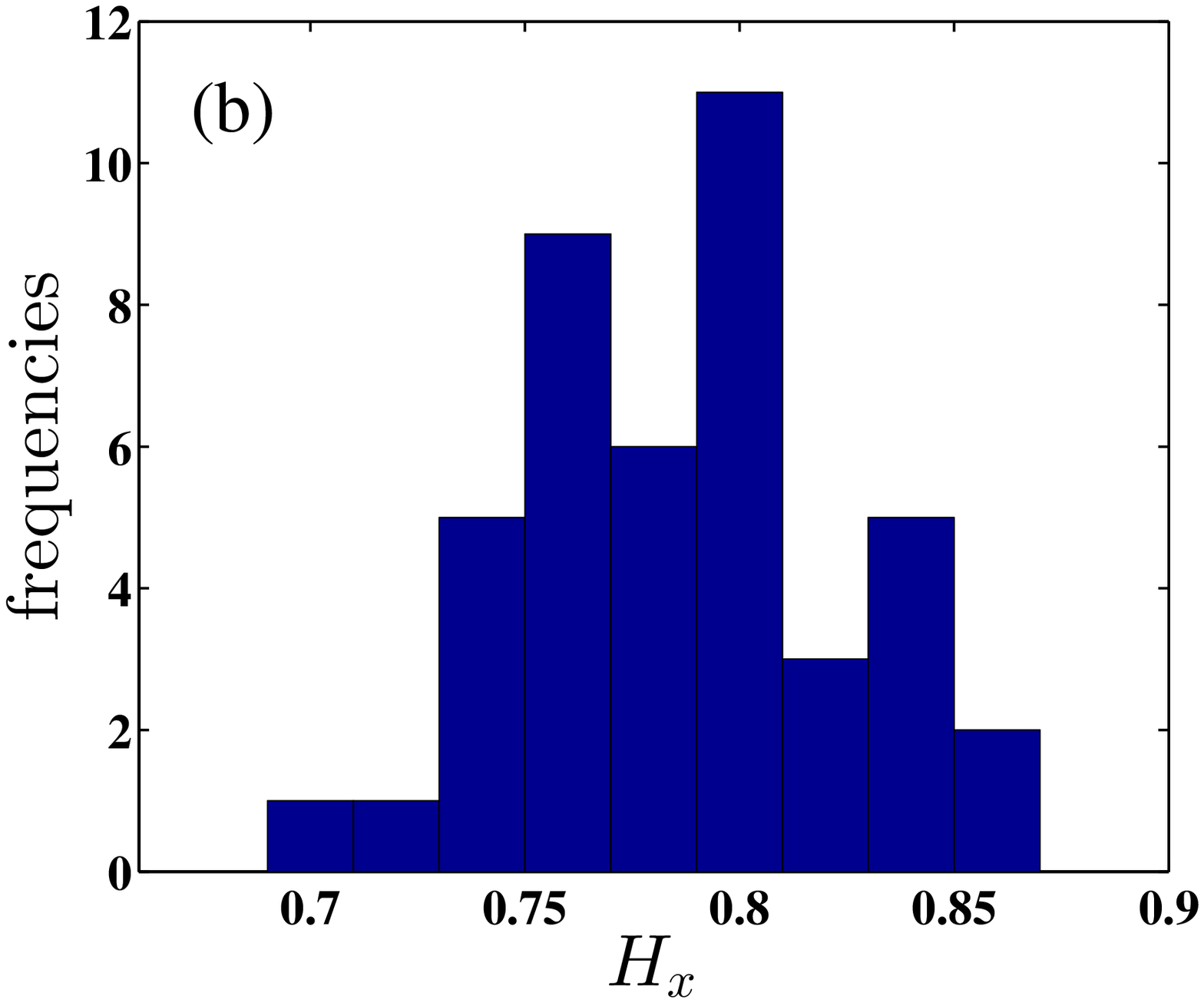}
  \caption{(colour online) Long memory in relative order prices. (a) Plots of the fluctuation functions $F(\ell)$ of relative order prices for four stocks 000001, 000839, 200488 and 200625. The solid lines are the least squares fits to the data. The scaling curves for 000839, 200488 and 200625 have been shifted vertically for clarity. (b) Histograms of Hurst exponents $H_x$ of relative order prices for all the 43 stocks.}
  \label{Fig:EBOD:Price:H}
\end{figure}

Memory effect of relative prices also plays an important role in model construction by introducing long memory in the volatility \citep{Gu-Zhou-2009-EPL,Zhou-Gu-Jiang-Xiong-Chen-Zhang-Zhou-2017-CE}. We study the memory effect using the DMA method and Fig.~\ref{Fig:EBOD:Price:H}(a) presents the fluctuation function $F(\ell)$ with respect to the scale size $\ell$ of relative prices for four representative stocks. Each curve reveals excellent power-law scaling behaviour with the scaling range spanning 4 to 5 orders of magnitude. Using the least-squares regression method, we obtain the Hurst exponents $H_x=0.847 \pm 0.009$ for stock 000001, $H_x=0.872 \pm 0.004$ for stock 000839, $H_x=0.736 \pm 0.005$ for stock 200488, and $H_x=0.763 \pm 0.006$ for stock 200625. The Hurst exponents of relative prices for all the 43 stocks are presented in Table~\ref{Tb:EBOD:Hurst:RelativePrice}. It is evident that the values are all significantly greater than 0.5. We conclude that the relative prices of placed orders have long memory. The histogram of the Hurst exponents of all stocks are showed in Fig.~\ref{Fig:EBOD:Price:H}(b), which confirms the long memory of relative prices in all stocks. According to Table~\ref{Tb:EBOD:Hurst:RelativePrice}, the average Hurst exponents are $\langle{H_x}\rangle=0.796 \pm 0.035$ for all stocks, $\langle{H_{x,A}}\rangle=0.808 \pm 0.030$ for A shares and $\langle{H_{x,B}}\rangle=0.761\pm 0.020$ for B-shares. It means that stronger persistence exists in relative prices of A shares in the year of 2003. We conjecture that the diagonal effect originally unveiled by \cite{Biais-Hillion-Spatt-1995-JF} is stronger for A-share stocks than for B-share stocks.

We propose a simple linear model between $H_x$ and $H_s$ and estimate the coefficients. It follows that
\begin{equation}
  H_x =0.471+0.466H_s
\end{equation}
in which both coefficients are significantly different from 0 at the 3\% significance level and the adjusted R-square is 0.179. Therefore, we can draw a consistent conclusion that stronger imitative and heading behaviour in a stock incurs stronger persistence in the order directions and relative prices.

\begin{table}[htp]
  \centering
  \caption{Hurst exponents $H_x$ of relative prices for the 32 A-share stocks and 11 B-share stocks estimated using the DMA method. The mean Hurst exponent is $\langle{H_x}\rangle=0.796 \pm 0.035$ for all stocks, $\langle{H_{x,A}}\rangle=0.8075 \pm 0.0301$ for A shares, and $\langle{H_{x,B}}\rangle=0.7606\pm 0.0204$ for B-shares.}
  \medskip
  \label{Tb:EBOD:Hurst:RelativePrice}
  \centering
  \begin{tabular}{cccc|cc|cc}
  \hline \hline
   Stock & $H_{x,A}$ & Stock & $H_{x,B}$  & Stock & $H_{x,A}$  & Stock & $H_{x,A}$  \\
  \hline
    000002 & 0.844 $\pm$ 0.005 & 200002 & 0.773 $\pm$ 0.002 & 000001 & 0.847 $\pm$ 0.009 & 000778 & 0.804 $\pm$ 0.007 \\
    000012 & 0.800 $\pm$ 0.005 & 200012 & 0.759 $\pm$ 0.004 & 000009 & 0.814 $\pm$ 0.011 & 000800 & 0.821 $\pm$ 0.008 \\
    000016 & 0.792 $\pm$ 0.005 & 200016 & 0.778 $\pm$ 0.003 & 000021 & 0.814 $\pm$ 0.008 & 000825 & 0.823 $\pm$ 0.004 \\
    000024 & 0.745 $\pm$ 0.007 & 200024 & 0.755 $\pm$ 0.002 & 000027 & 0.802 $\pm$ 0.012 & 000839 & 0.872 $\pm$ 0.004 \\
    000429 & 0.767 $\pm$ 0.008 & 200429 & 0.720 $\pm$ 0.005 & 000063 & 0.840 $\pm$ 0.019 & 000858 & 0.801 $\pm$ 0.005 \\
    000488 & 0.792 $\pm$ 0.002 & 200488 & 0.736 $\pm$ 0.005 & 000066 & 0.808 $\pm$ 0.006 & 000898 & 0.847 $\pm$ 0.007 \\
    000539 & 0.800 $\pm$ 0.004 & 200539 & 0.741 $\pm$ 0.008 & 000088 & 0.776 $\pm$ 0.005 & 000917 & 0.766 $\pm$ 0.008 \\
    000541 & 0.765 $\pm$ 0.004 & 200541 & 0.776 $\pm$ 0.002 & 000089 & 0.756 $\pm$ 0.008 & 000932 & 0.800 $\pm$ 0.009 \\
    000550 & 0.828 $\pm$ 0.004 & 200550 & 0.775 $\pm$ 0.004 & 000406 & 0.787 $\pm$ 0.005 & 000956 & 0.861 $\pm$ 0.006 \\
    000581 & 0.819 $\pm$ 0.002 & 200581 & 0.793 $\pm$ 0.002 & 000709 & 0.807 $\pm$ 0.005 & 000983 & 0.789 $\pm$ 0.004 \\
    000625 & 0.813 $\pm$ 0.005 & 200625 & 0.763 $\pm$ 0.006 & 000720 & 0.841 $\pm$ 0.003 &        &                   \\
  \hline\hline
 \end{tabular}
\end{table}

\subsection{Order size}

The last ingredient is the order size $v$. In the model, we do not simply study the statistical properties of order size, but analyze the relation between the order size $v$ and relative price $x$. It is due to the fact that there is a remarkable number preference in the distribution of order size \citep{Mu-Chen-Kertesz-Zhou-2009-EPJB}, which suggests that the distribution is singular almost everywhere and cannot be modelled in a feasible way.

\begin{figure}[htb]
  \centering
  \includegraphics[width=7.5cm,height=6cm]{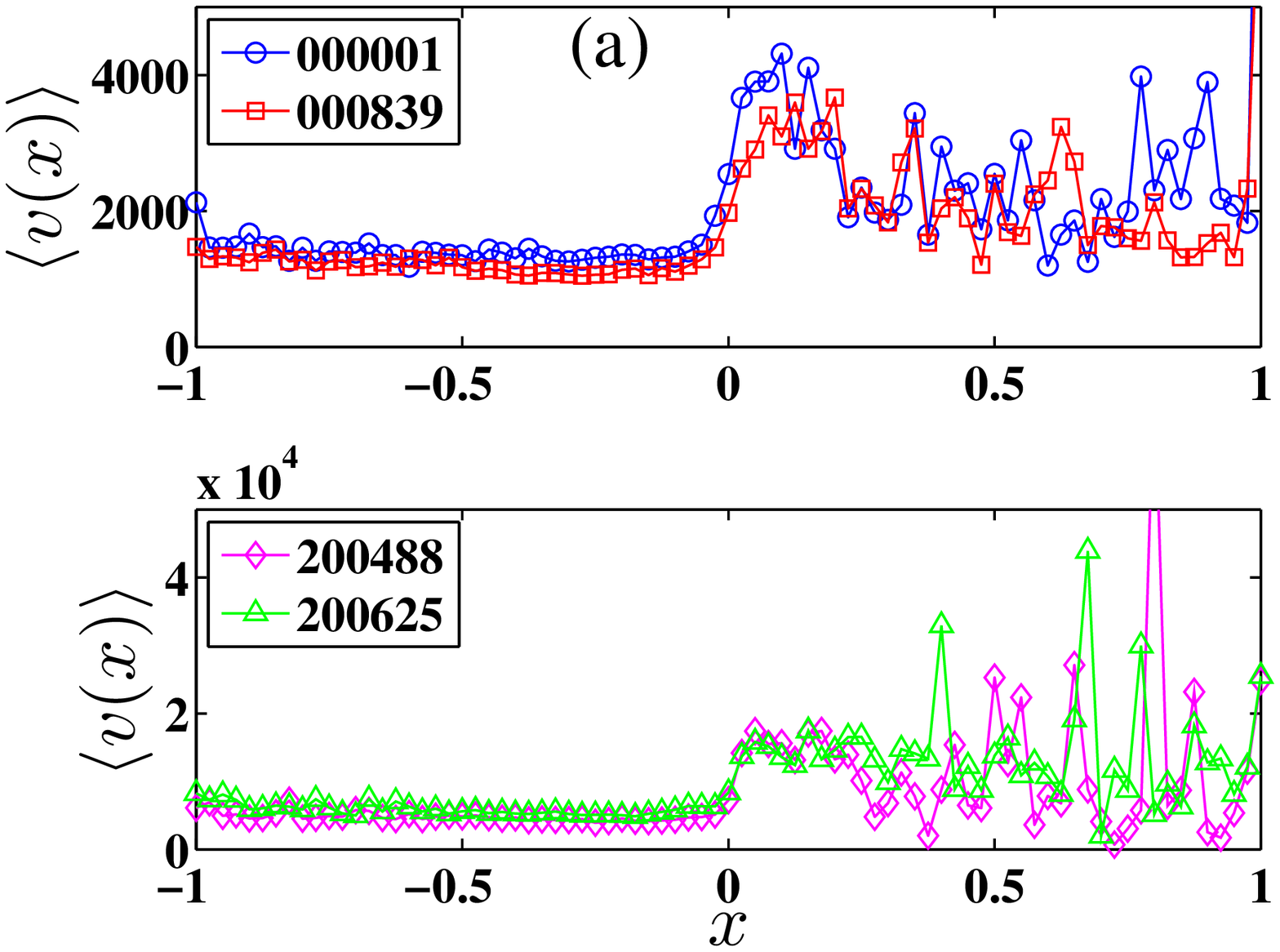}\hspace{3mm}
  \includegraphics[width=7.5cm,height=6cm]{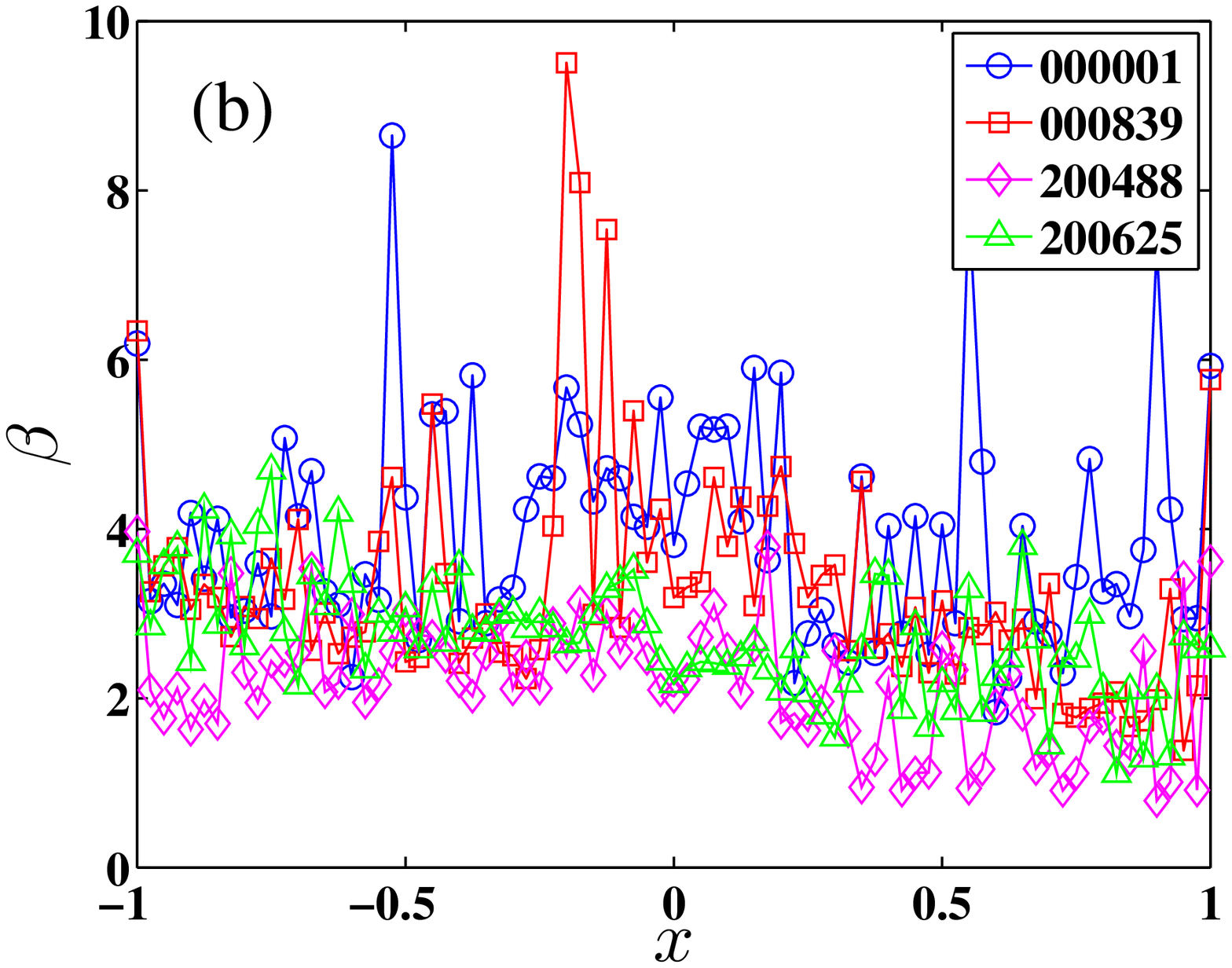}
  \caption{(colour online) Dependence of order size with respect to relative price. (a) The mean sizes $\langle{v(x)}\rangle$ against the relative prices $x$ of empirical data for four stocks 000001, 000839, 200488 and 200625. (b) The ratio $\beta(x)$ as a function of $x$ for the same four stocks.}
  \label{Fig:EBOD:OrderSize}
\end{figure}

The relationship between order sizes and relative prices of the same four representative stocks is presented in Fig.~\ref{Fig:EBOD:OrderSize}(a). For each stock, we divided the submitted orders into many groups by binning their relative prices, and the average relative price and average order size are calculated for each group of orders. We find that these curves have similar shapes. Overall, market orders have larger sizes than limit orders. We observe that the average order size $\langle{v(x)}\rangle$ is almost independent of the relative price for effective limit orders ($x<0$). There is an interesting feature showing that around the point $x=0$ the average size $\langle{v(x)}\rangle$ increases rapidly with the relative price $x$, which indicates that traders tend to place larger orders around the opposite price level. For market orders with large $x$ values, the average order size fluctuates a lot. It is because that the number of such market orders is relatively small. Impatient traders submit effective market orders with large relative prices to ensuring the execution. However, most orders do not penetrate many levels on the opposite book. Moreover, we find that the order sizes of A-share stocks (000001 and 000839) are smaller than the ones of B-share stocks (200488 and 200625). This result is also confirmed by other A shares and B shares stocks, indicating that investors submitted larger orders in the B-share markets, which is because the proportion of retailer traders is much higher in the A-share market.

We assume that the distribution of $x$ is normal for a given $x$. We calculate the ratio $\beta$ which is the standard deviation of order sizes $\sigma(v)$ divided by the average size $\langle{v(x)}\rangle$
\begin{equation}
  \beta=\sigma(v)/\langle{v(x)}\rangle.
  \label{Eq:EBOD:beta}
\end{equation}
Fig.~\ref{Fig:EBOD:OrderSize}(b) illustrates the relation between the ratio $\beta$ and the relative price $x$ for the four stocks. We find that the ratio $\beta$ almost fluctuates around a constant value for the four stocks.

\subsection{Order cancellation}

Order cancellation is another main process of continuous double auction. It refers to removing stale orders from the LOB. Order cancellation plays an important role in price formation of security markets. If all the orders placed at the best price are cancelled, the mid-price defined as the mean value of the best bid and best ask will change. If a cancellation takes place inside the LOB, it has a potential impact on price movement. The model of \cite{Mike-Farmer-2008-JEDC} considers three factors in the conditional order cancellation probability, i.e., the relative distance of the target order to the opposite best, the ratio of buy or sell orders on the book and the total number of LOB orders. However, these factors are not observable to the traders in the Chinese market. In addition, \cite{Gu-Zhou-2009-EPL} find that the main stylized facts can be reproduced if one uses a Poisson process for order cancellation. What is important in the LOB dynamics is the rate of cancellation, together with the rates of limit order placement and market order placement. Hence, we adopt a probability cancellation method in the model, as first reported by \cite{Gu-Xiong-Ren-Zhou-Zhang-2013-JSM}, which can capture the proper rate of order cancellation.

\begin{figure}[htb]
  \centering
  \includegraphics[width=7.5cm,height=6cm]{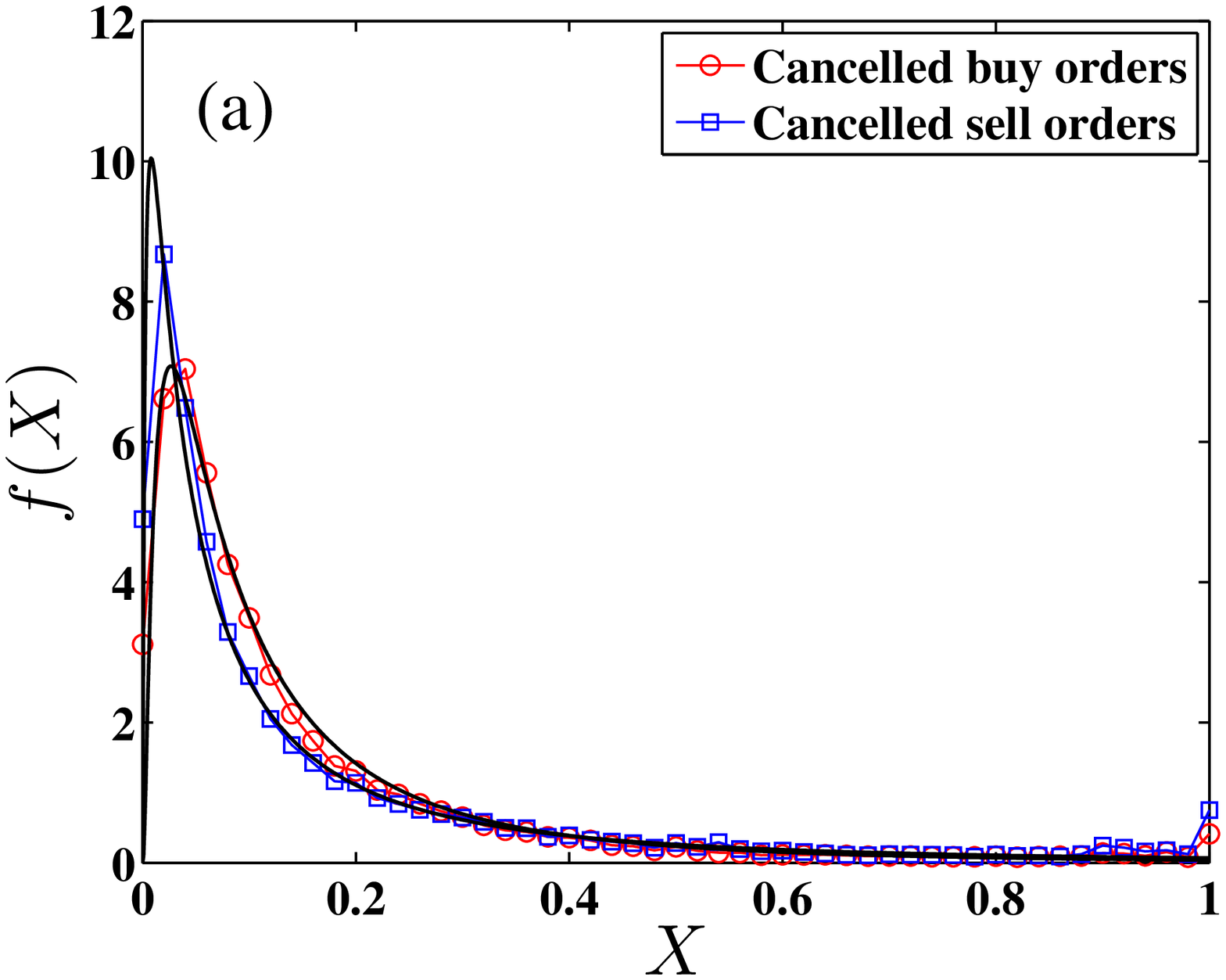}
  \includegraphics[width=7.5cm,height=6cm]{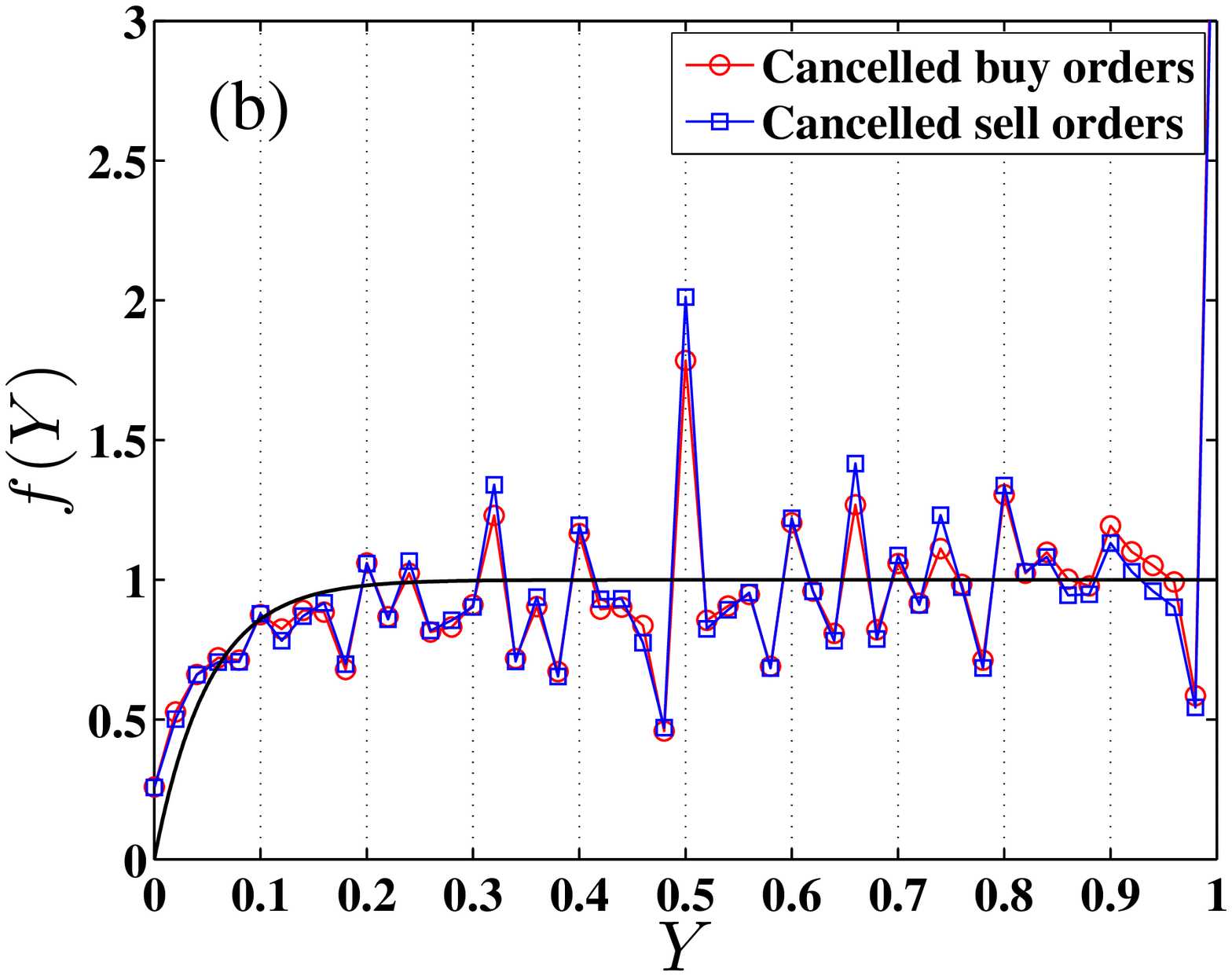}
  \caption{(colour online) Determination of the order cancellation process. (a) Probability density functions $f(X)$ of relative price levels on the LOB for both cancelled buy and sell orders of stock 000001. The solid lines are fits to the rescaled log-normally distribution. (b) Probability density functions $f(Y)$ of relative temporal positions at all price levels for both cancelled buy and sell orders of the same stock. The solid line is the fit to an exponential function.}
  \label{Fig:EBOD:Cancellation}
\end{figure}

The position of an order on the LOB at time $t$ is fully determined by its price level or spatial position $l(t)$ (space dimension) and its temporal position $y(l,t)$ (time dimension) in the order queue at the $l$-th level \citep[Fig.1]{Gu-Xiong-Ren-Zhou-Zhang-2013-JSM}. First, we define the relative price levels $X_i$,
\begin{equation}
 X(t)=\frac{l_t}{L_{b,s}(t)}~,
 \label{Eq:X}
\end{equation}
where $l(t)$ is the price level in which a cancellation occurs in the LOB at event time $t$ and $L_{b,s}(t)$ is the total number of price levels existing in the buy or sell LOB. The relative level $X$ varies in the range $\left(0,1\right]$. A small value of $X$ refers to a cancellation happening close to the same best, while a large value of $X$ means that a cancellation occurs far from the same best. Fig.~\ref{Fig:EBOD:Cancellation}(a) presents the PDF $f(X)$ for both cancelled buy and sell orders of stock 000001.
As shown by \cite{Gu-Xiong-Ren-Zhou-Zhang-2013-JSM}, the PDF $f(X)$ of relative price levels follows a rescaled log-normal distribution
\begin{equation}
 f(X)=\frac{1}{z}\frac{1}{\sqrt{2\pi}\sigma{X}}\exp\left[-\frac{(\ln{X}-\mu)^2}{2\sigma^2}\right]~,
 \label{Eq:pdf-X}
\end{equation}
where $\mu$ is the location parameter, $\sigma$ is the scaling parameter, and $z$ is the normalization constant ensuring that $\int_0^1{f(X)}dX=1$. Using the least-squares fitting method, we obtain $\mu=-2.36$ and $\sigma=1.13$ for cancelled buy orders and $\mu=-2.49$ and $\sigma=1.52$ for cancelled sell orders.

After the price level $l$ of the cancellation is determined, we need to determine which order will be cancelled at the price level. Denote $y(l,t)$ as the temporal position of a cancelled order in the queue of the $l$-th price level at time $t$. An order with $y(l,t)=1$ is the order placed the earliest in the queue. In order to removing the number impact of orders stored at the $l$-th price level, we analyze the relative temporal position $Y(l,t)$ instead of $y(l,t)$,
\begin{equation}
 Y(l,t)=\frac{y(l,t)}{N_{b,s}(l,t)}~,
 \label{Eq:Y}
\end{equation}
where $Y(l,t)$ varies in the range $(0,1]$ and $N_{b,s}(l,t)$ is the total number of orders stored at the $l$-th price level on the buy or sell LOB at time $t$. Since the PDF $f(Y)$ at each price level in the LOB has similar shape, we aggregate the data $Y(x,t)$ at all price levels together and treat them as an ensemble. The ensemble PDFs $f(Y)$ for both cancelled buy and sell orders of stock 000001 are presented in Fig.~\ref{Fig:EBOD:Cancellation}(b). The function $f(Y)$ is close to zero when the relative temporal position $Y$ approaches to zero. When $Y$ increases, the PDF first increases rapidly in the range $Y\leq0.1$ and then fluctuates around a constant level until the end of the LOB. These observations indicate an interesting feature that patient traders have better self-discipline since they place their orders early at certain levels and are less prone to cancel the orders. The $f(Y)$ value is extremely larger when $Y=1$, which indicates that the latest placed orders are more likely to be cancelled. We can apply an exponential function to fit the PDF of relative temporal position $Y$ as follows,
\begin{equation}
 f(Y)=\frac{1}{z}(1-e^{{\gamma}Y})~,
 \label{Eq:pdf-Y}
\end{equation}
where $\gamma$ is the exponent and $z=(\gamma+1-e^\gamma)/\gamma$ is the normalization factor. Using the least-squares fitting method, we obtain $\gamma=-33.78$ for cancelled buy orders and $\gamma=-36.57$ for cancelled sell orders.

\subsection{Price formation}

Assume that there are $N(T)$ steps of order placements on a trading day $T$. The price formation process is carried out as follows.

We first generate an order direction (sign) series $\{s_i: i=1,2,\cdots,N(T)\}$ containing just two elements ``$+1$'' for buy orders and ``$-1$'' for sell orders with the memory effect characterized by the Hurst exponent $H_s$ from real data. There are many different algorithms for the generation of fractional Brownian motions (FBMs) \citep{Bardet-Lang-Oppenheim-Philippe-Stoev-Taqqu-2003}. We adopt the wavelet-based algorithm to generate FBMs of size $N(T)+1$ \citep{Abry-Sellan-1996-ACHA}, which is an excellent FBM generator especially for small Hurst exponents  \citep{Ni-Jiang-Zhou-2009-PLA,Shao-Gu-Jiang-Zhou-Sornette-2012-SR,Shao-Gu-Jiang-Zhou-2015-Fractals,Qian-Liu-Jiang-Podobnik-Zhou-Stanley-2015-PRE}. The sign sequence of the increments of the generated FBM is assigned to $\{s_i\}$.

We then generate a relative price series $\{x_i: i=1,2,\cdots,N(T)\}$ with given degree of long memory quantified by the Hurst index $H_x$ \citep{Gu-Zhou-2009-EPL}. A sequence of relative prices $\{x_{i,0}: i=1,2,\cdots,N(T)\}$ are drawn from the probability distribution of real data presented in Fig.~\ref{Fig:EBOD:Price:PDF}, in which $x_{i,0}$ is obtained by solving $z_i = F(x_{i,0})$, where $\{z_i: i=1,2,\cdots,N(T)\}$ is a sequence of random numbers drawn from the uniform distribution defined in $[0,1]$ and $F(x)$ is the cumulative distribution of $f(x)$ \citep{Press-Teukolsky-Vetterling-Flannery-1996}. To introduce a memory effect to the relative price series $\{x_0\}$ with the Hurst exponent $H_x$, we simulate a FBM with $H_x$ and record its increments as $\{y_i: i=1, 2, \cdots, N(T)\}$. The sequence $\{x_{i,0}: i=1,2,\cdots, N(T)\}$ is rearranged such that $x_i$ ranks the $n$-th in sequence $\{x_i: i=1,2,\cdots, N(T)\}$ if and only if $y_i$ ranks the $n$-th in the $\{y_i: i=1,2,\cdots, N(T)\}$ sequence \citep{Bogachev-Eichner-Bunde-2007-PRL,Zhou-2008-PRE}. A detrending moving average analysis of $x_i$ confirms that its DMA scaling exponent is very close to $H_x$.

Finally, we generate the order size sequence $\{v_i: i=1,2,\cdots,N(T)\}$ from a Normal distribution with the average order size $\langle{v(x_i)}\rangle$ and its standard deviation $\beta(x_i)x_i$, where, for each $x_i$, the corresponding $\langle{v(x_i)}\rangle$ and $\beta(x_i)$ are determined according to their relations to the relative price depicted in Fig.~\ref{Fig:EBOD:OrderSize}.

Having generated three ingredients $\{s_i\}$, $\{x_i\}$ and $\{v_i\}$ of an order, we continue performing the price formation process. For an order placed at the $t$-th step ($t=1,2,\cdots,N(T)$), we transform Eq.~(\ref{Eq:EBOD:x}) for buy orders ($s_i=+1$) and sell orders ($s_i=-1$) to obtain the order price $p_i$, which is rounded to two decimals since the tick size is 0.01 yuan in the Chinese stock market. The $t$-th order is compared with the current buy or sell LOB to determine whether it is fully or partly executed or stored in the LOB according to the price-time priority mechanism of continuous double auction. If a transaction occurs, the mid-price $m(t)$ defined as the mean value of the best bid and best ask is recorded.

At each step, we check whether a cancellation occurs according to a Poisson process with its characteristic parameter obtained from the real data (it is 19\% for stock 000001). If a cancellation occurs, we first determine the price level $X$ of cancellation location in the LOB based on the log-normal distribution presented in of Fig.~\ref{Fig:EBOD:Cancellation}(a), and then obtain the cancellation position $Y$ at the price level $X$ according to the exponential distribution illustrated in Fig.~\ref{Fig:EBOD:Cancellation}(b).

In order to avoid the buy or sell LOB becoming too empty, one needs to impose an additional regulation that there are at least two orders at each side of the LOB after a transaction or cancellation \citep{Mike-Farmer-2008-JEDC,Gu-Zhou-2009-EPL}.

After simulating $N(T)$ steps of order placements, the price formation process is completed on the $T$-th trading day. We can start a new simulation on the $(T+1)$-th trading day with the same process. We mention that on the $(T+1)$-th trading day the maximum price $p_{\rm{max}}(T+1)=(1+\phi_+)p_{\rm{c}}(T)$ and the minimum price $p_{\rm{min}}(T+1)=(1+\phi_-)p_{\rm{c}}(T)$, where $p_{\rm{c}}(T)$ is the closing price on trading day $T$, which is the average of the last 100 values of the simulated mid-prices $m(j)$. According to the $\pm10\%$ price limit rule in the Chinese stock market, we use $\phi_+=0.10$ and $\phi_-=-0.10$ for model validation.

\subsection{Model validation}

When a model is built, one needs to calibrate it with known stylized facts \citep{Li-Zhang-Zhang-Zhang-Xiong-2014-IS}. The most universal stylized facts in stock markets are the absence of autocorrelations in the returns, power-law tails in the return distribution and long memory in the volatility time series \citep{Cont-2001-QF}. The modified Mike-Farmer model of \cite{Gu-Zhou-2009-EPL} can well reproduce these three stylized facts. We illustrate briefly the model validation of these three stylized facts.

We simulate more than two hundred trading days of price formation and obtain a mid-price time series $m(t)$. In the simulations, we set set $\phi_+=0.10$ and $\phi_-=-0.10$ as the real stocks. The returns are calculated as the logarithmic differences of mid-price,
\begin{equation}
 R(t)=\ln[m(t)/m(t-1)],
 \label{Eq:R}
\end{equation}
and the volatility is defined as the absolute value of return,
\begin{equation}
 V_i=|R_i|.
 \label{Eq:V}
\end{equation}

\begin{figure}[htb]
\centering
  \includegraphics[width=5cm,height=4cm]{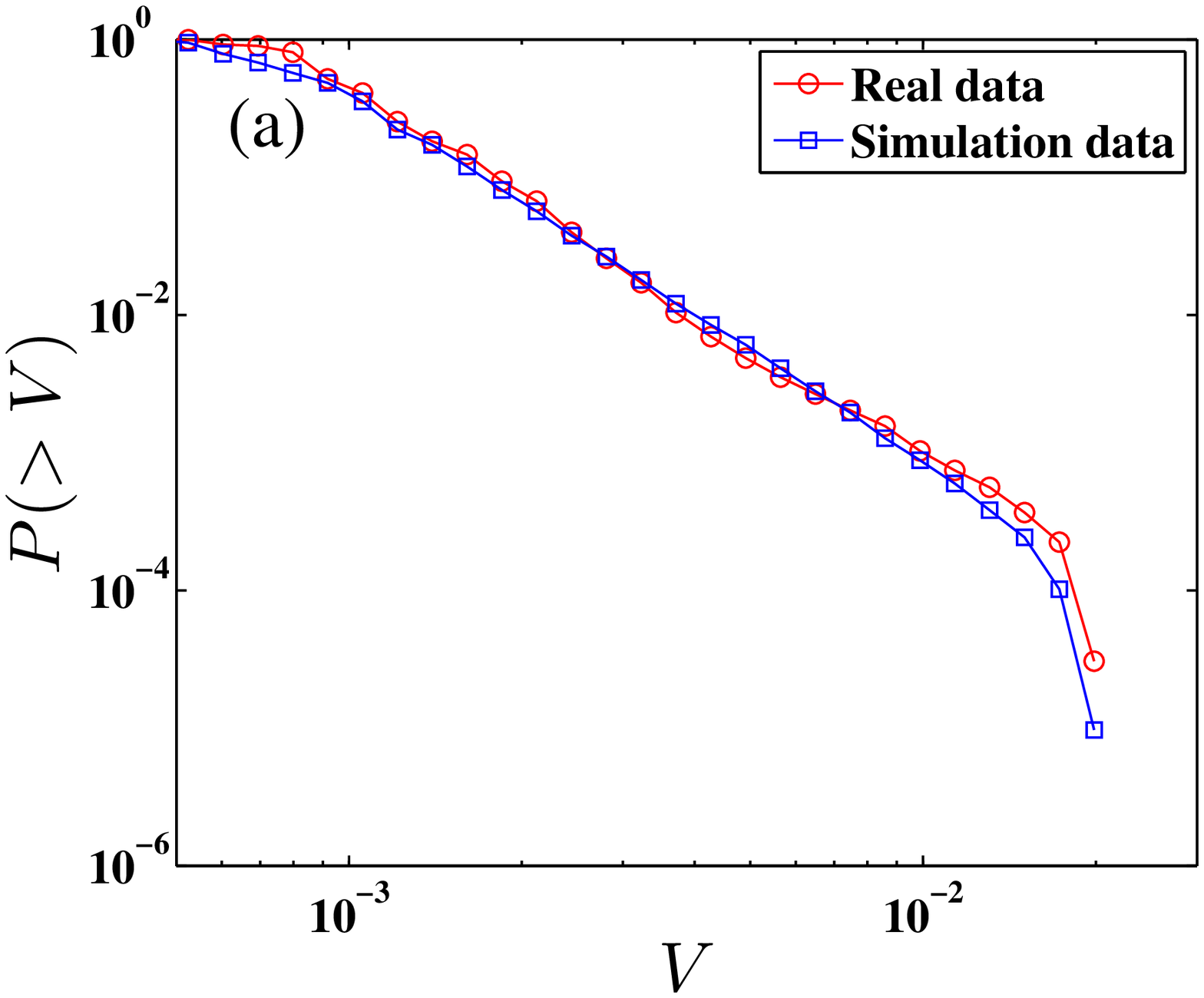}
  \includegraphics[width=5cm,height=4cm]{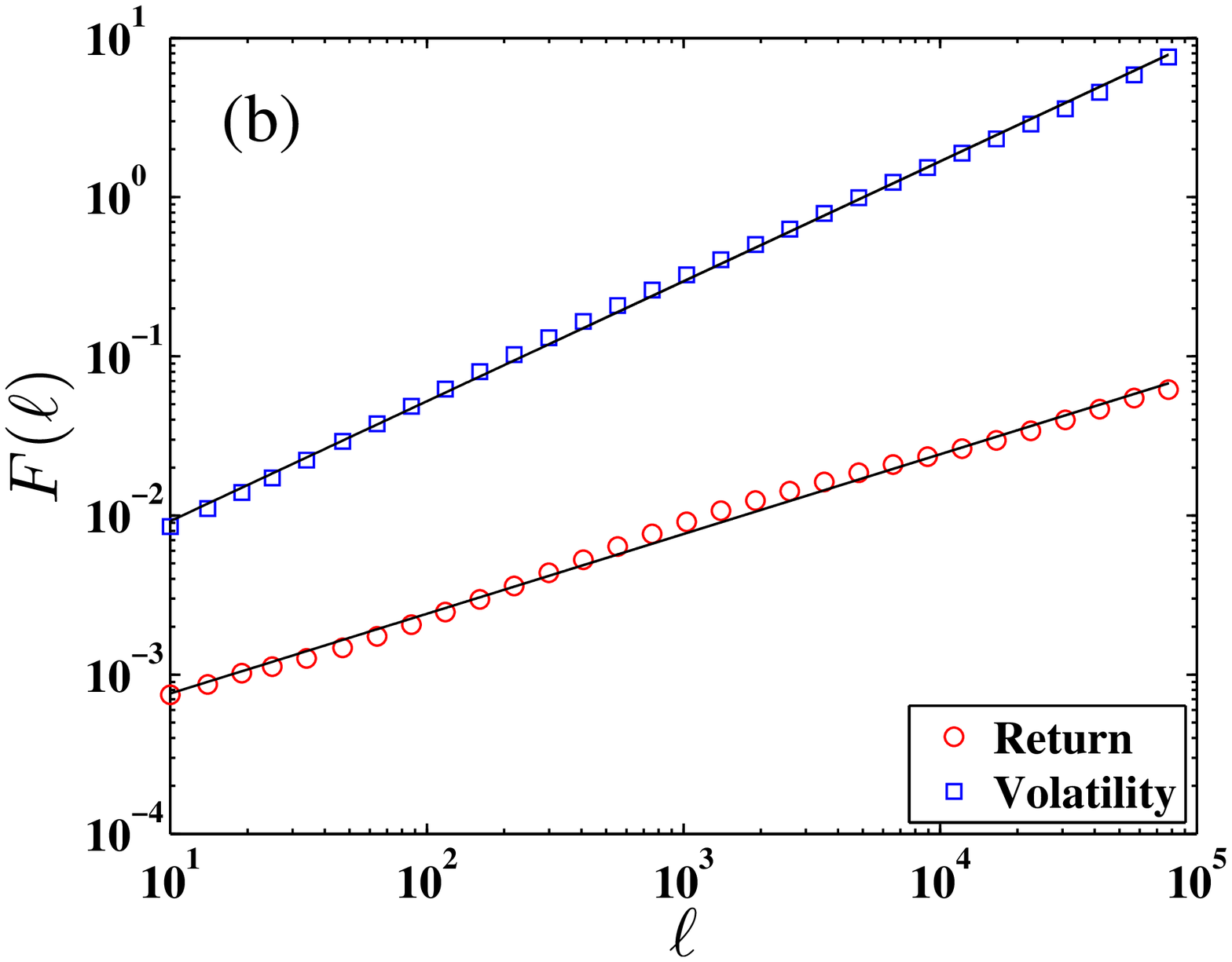}
  \includegraphics[width=5cm,height=4cm]{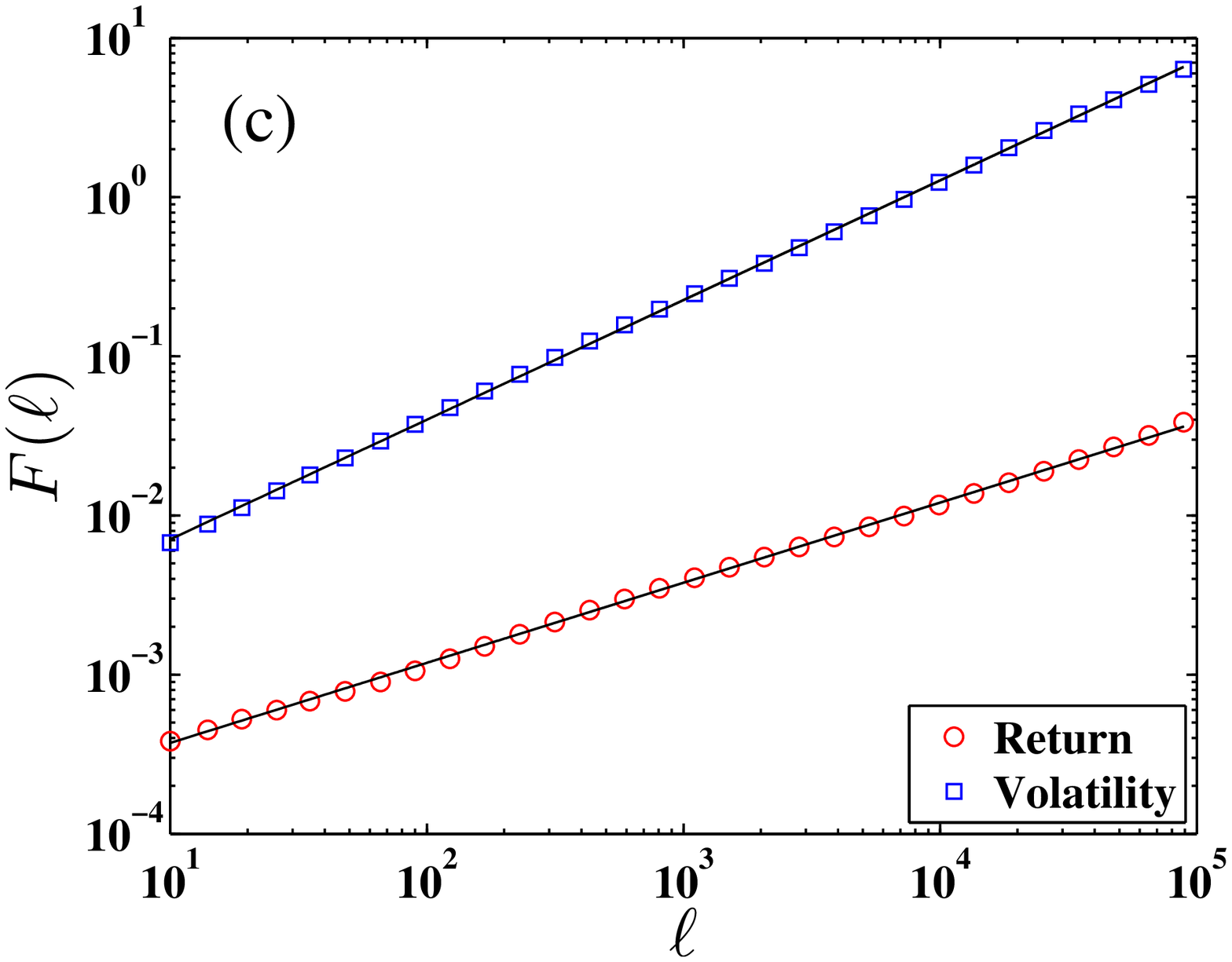}
  \caption{(color online) Model validation. (a) Empirical cumulative distributions of volatility of simulated data and stock 000001. Both distributions have power-law tails. (b) DMA fluctuation functions $F(\ell)$ of the return and volatility series for simulation data. (c) DMA fluctuation functions $F(\ell)$ of the return and volatility series of Stock 000001. The curves of volatility in (b) and (c) have been shifted vertically for clarity.}
\label{Fig:EBOD:Validation}
\end{figure}

Fig.~\ref{Fig:EBOD:Validation}(a) presents the cumulative volatility distributions of the simulated data from the order-driven model and the real data of stock 000001. The two distributions overlap with each other and have power-law tails. Using an efficient quantitative method proposed by \cite{Clauset-Shalizi-Newman-2009-SIAMR}, which is based on the Kolmogorov-Smirnov test and the maximum likelihood estimation (MLE) method, we confirm the presence of power-law tails
\begin{equation}
  P(>V) \sim V^{-\alpha}
  \label{Eq:EBOD:PDF:r}
\end{equation}
and obtain the tail exponent $\alpha=2.65\pm0.02$ for the model and $\alpha=2.96\pm0.03$ for the stock. The results comply with the empirical findings of Chinese stocks \citep{Gu-Chen-Zhou-2008a-PA}. It shows that the tail distribution of volatility obeys the universal inverse cubic law \citep{Gopikrishnan-Meyer-Amaral-Stanley-1998-EPJB}. The discrepancy of tail exponent between the simulated and real stock returns comes from the fact that the method of \cite{Clauset-Shalizi-Newman-2009-SIAMR} gives $V_{\min}=   0.0017$ for the stimulated returns and $V_{\min} = 0.0024$ for the real data in which the tail exponents are estimated from the absolute returns that is no less than $V_{\min}$.

We then apply the DMA method to compare the memory effects of simulated data with real data of stock 000001. Fig.~\ref{Fig:EBOD:Validation}(b) shows the fluctuation functions $F(\ell)$ with respect to the size scales $\ell$ of simulated return and volatility series. It is clear that the function $F(\ell)$ scales with the scale $\ell$ as a power law for both curves. Using the least squares fitting method, we obtain the DMA scaling exponents $H_R=0.501\pm0.007$ for the return series and $H_V=0.753\pm0.005$ for the volatility series of the simulation data. In Fig.~\ref{Fig:EBOD:Validation}(c), we show the fluctuation functions $F(\ell)$ of return and volatility series for the real data and find that $F(\ell)$ also scales with $\ell$ as a power law. We obtain that $H_R=0.503\pm0.003$ for the return series and $H_V=0.752\pm0.002$ for the volatility series. The results of simulated returns and volatilities are consistent with stylized facts that return time series is uncorrelated with the Hurst exponent close 0.5 whereas the volatility series processes long memory with the Hurst exponent significantly greater than 0.5.

\section{Computational experiments on the effects of asymmetric price limits}
\label{sec:meodelapp}

Motivated by the debate on the {\textit{Draft Guideline on the Trading of Stocks Bearing Risk Warnings}}, we perform computational experiments to investigate the effects of asymmetric price limit mechanism on the price dynamics and some stylized facts. In the computational experiments, we change the value of $\phi_+$ from 0.05 to 0.3 with the increment of 0.05 and the value of $\phi_-$ from -0.3 to -0.05 with the same increment. The initial price is 10 yuan. The tick size is 0.01 yuan.

\subsection{Vanishing and divergence of price trajectory}

\begin{figure}[htb]
  \centering
  \includegraphics[width=5cm]{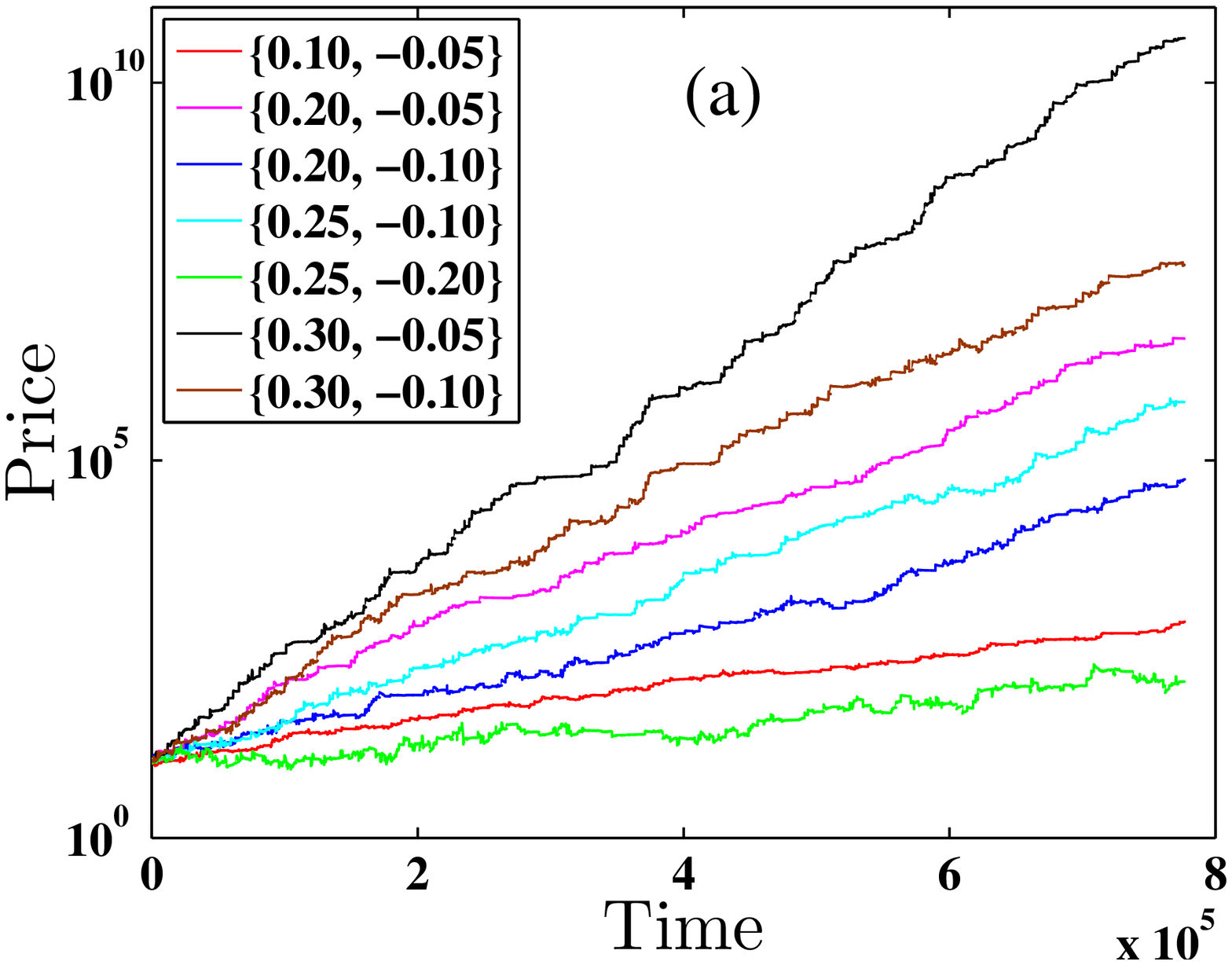}
  \includegraphics[width=5cm]{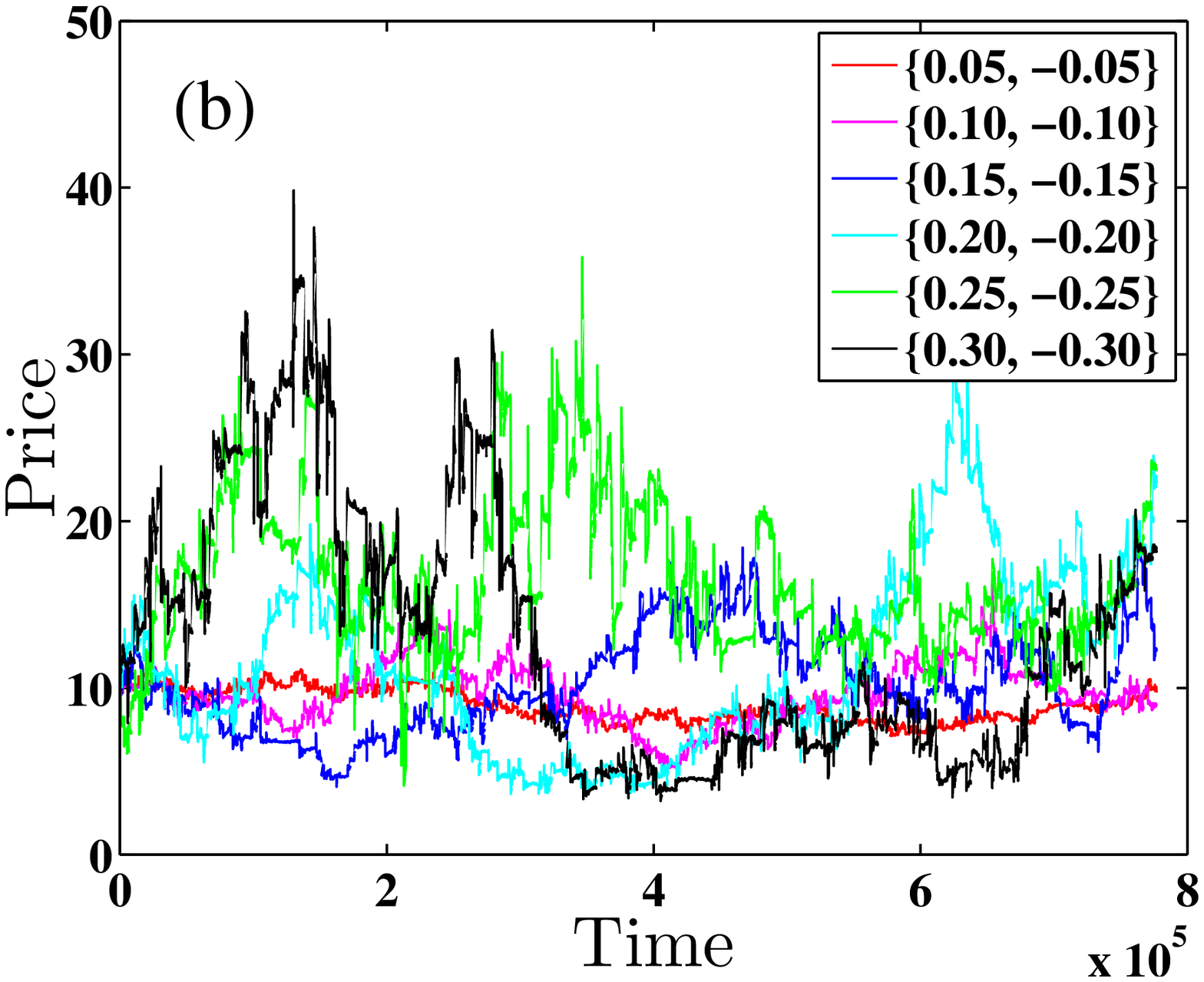}
  \includegraphics[width=5cm]{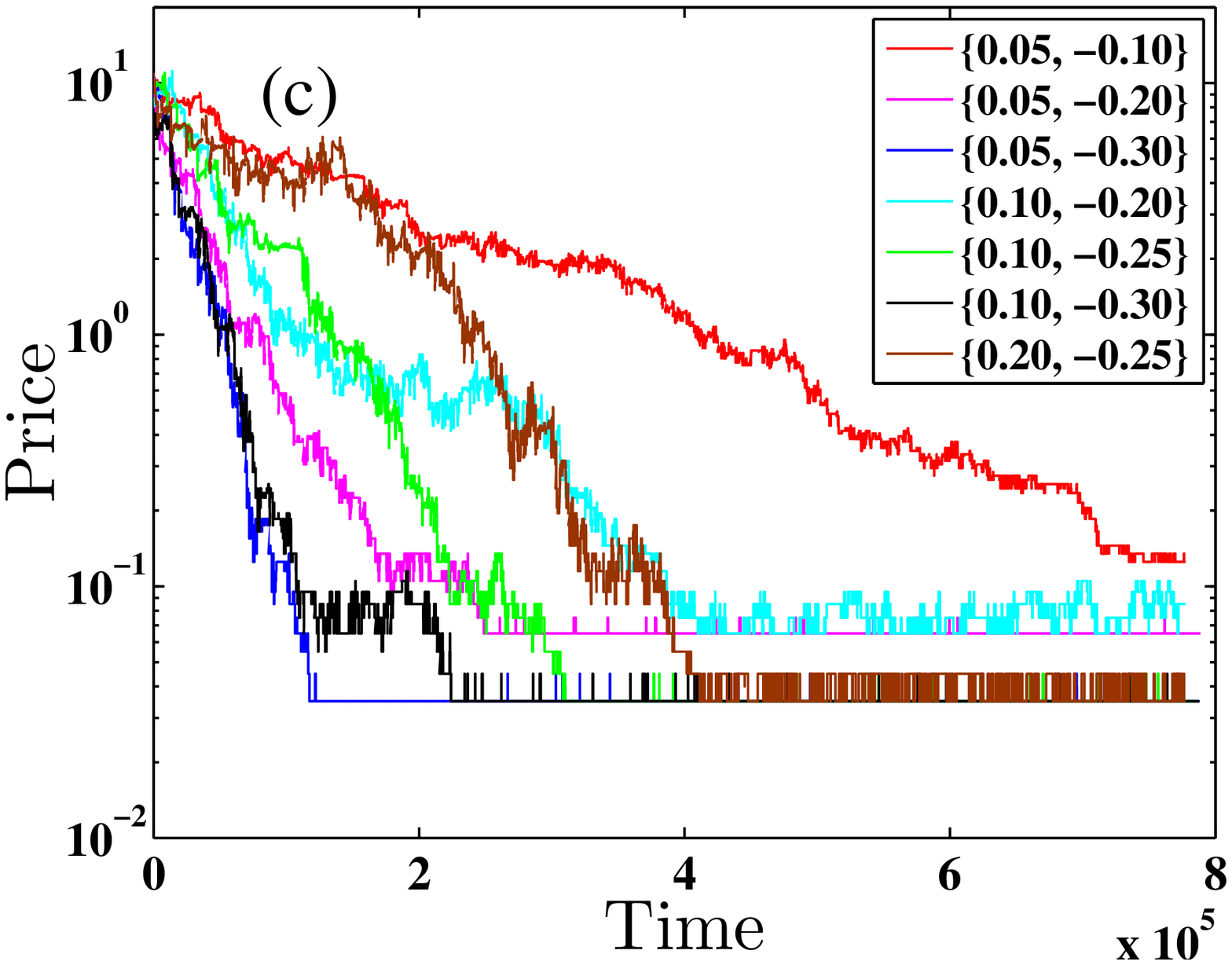}
  \caption{(color online) Price trajectories of the stocks simulated for different combinations of price limits $\{\phi_+,\phi_-\}$. (a) $|\phi_-|<\phi_+$. (b) $|\phi_-|=\phi_+$. (c) $|\phi_-|>\phi_+$. }
\label{Fig:EBOD:Price:Sim}
\end{figure}

Fig.~\ref{Fig:EBOD:Price:Sim} presents the evolution of simulated mid-price for different combinations of $\{\phi_+,\phi_-\}$: $|\phi_-|<\phi_+$ in panel (a), $\phi_+=|\phi_-|$ in panel (b), and $|\phi_-|>\phi_+$ in panel (c). It is evidence that the patterns of the price trajectories are similar in each plot but completely different among different plots.

When $|\phi_-|<\phi_+$, the price will go up quickly to an unreasonable extent and diverge, as shown in Fig.~\ref{Fig:EBOD:Price:Sim}(a). The price diverges faster if the up limit is larger or the absolute down limit is smaller. Let us take a buy order for example. We can obtain the following equation from Eq.~(\ref{Eq:EBOD:x}),
\begin{equation}
p_i = \left\{
\begin{array}{ll}
    x[p_{\rm{max}}(T) - p_a(t-1)] + p_a(t-1)&x \geq 0 \\
    x[p_a(t-1) - p_{\rm{min}}(T)] + p_a(t-1)&x < 0
\end{array}
\right..
 \label{Eq:p_buy}
\end{equation}
We know that $p_i$ is always proportional to $p_{\rm{max}}$ and $p_{\rm{min}}$. When an efficient buy market order ($x \geq 0$) is placed, the price $p_i$ increases with the $p_{\rm{max}}$, which means that the buy price is very aggressive and it tends to push up the stock price when $p_{\rm{max}}$ increases. On the other hand, if a trader places an efficient buy limit order ($x < 0$), $p_i$ also increases with the $p_{\rm{min}}$. It means that the trader places the limit order with higher price in the limit order book when $p_{\rm{min}}$ increases, which makes the stock price grow as well. The explanation for sell orders is similar. Hence, the average return is greater than 0.

When $|\phi_-|=\phi_+$, the price evolves in a reasonable range, as shown in Fig.~\ref{Fig:EBOD:Price:Sim}(b). With the increase of $-\phi_-$ and $\phi_+$, the fluctuations of price enhance. We find that there are sharp rises and drops, which are actually bubbles and crashes frequently observed in the Chinese stock market \citep{Zhou-Sornette-2004a-PA,Jiang-Zhou-Sornette-Woodard-Bastiaensen-Cauwels-2010-JEBO}. This observation is very interesting, because it suggests that these artificial traders can also trigger collective behaviours. The model is thus not a zero-intelligence model. Traders' strategies and traits have been well captured by the micro regularities used in this model.

When $|\phi_-|>\phi_+$, the price drops very quickly, as shown in Fig.~\ref{Fig:EBOD:Price:Sim}(c). The price decays faster if the up limit is smaller or the absolute down limit is larger. For each curve, there seems to have a lower bound ${\cal{B}}$ such that the price cannot be lower. The presence of this lower bound is actually caused by the presence of down limit and tick size and the formation rule of closing prices. Assuming that the closing price on day $T-1$ is ${\cal{B}}$, the minimum price on day $T$ is $\prec{\cal{B}}(1-|\phi_-|)\succ$, where $\prec\succ$ is the round operator. We note that
\begin{equation}
  {\cal{B}}=\max\{{\cal{B}}: 100{\cal{B}}-\prec{100\cal{B}}(1-|\phi_-|)\succ = 1\}.
  \label{Eq:EBOD:LowerBound:Cond}
\end{equation}
In this way, together with the tick size condition
\begin{equation}
    \prec100{\cal{B}}\succ=100{\cal{B}},
 \label{Eq:EBOD:LowerBound:Ticksize}
\end{equation}
the minimum closing price will be
\begin{equation}
  \prec100{\cal{B}}/2+\prec{100\cal{B}}(1-|\phi_-|)\succ/2\succ = \prec100{\cal{B}}+0.5\succ = 100{\cal{B}}.
\end{equation}
In order words, the lower bound of the prices is reached at ${\cal{B}}$. It follows from Eq.~(\ref{Eq:EBOD:LowerBound:Cond}) that
\begin{equation}
  {\cal{B}}=\max\{{\cal{B}}: 0.5 < 100{\cal{B}}|\phi_-| \le 1.5\},
  \label{Eq:EBOD:LowerBound:Cond:bias}
\end{equation}
or
\begin{equation}
  {\cal{B}}=\max\left\{{\cal{B}}: \frac{0.005}{|\phi_-|} < {\cal{B}} \le \frac{0.015}{|\phi_-|}\right\}.
  \label{Eq:EBOD:LowerBound}
\end{equation}
Therefore, for $\phi_-=-0.05$, -0.10, -0.15, -0.20, -0.25 and -0.30, we obtain that ${\cal{B}}=0.30$, 0.15, 0.10, 0.07, 0.06 and 0.05. These values fit the empirical results in Fig.~\ref{Fig:EBOD:Price:Sim}(c) very well.

Fig.~\ref{Fig:EBOD:Price:Sim}(a) shows that, when $|\phi_-|<\phi_+$, the price increases exponentially
\begin{equation}
  p_t \sim e^{\lambda{t}}.
\end{equation}
We introduce the divergence rate $\lambda$ of the price to quantify the impacts of these asymmetric price limit rules. The divergence rates of all cases are estimated, which are presented in the upper triangle of Table \ref{Tb:EBOD:ComputExper:Price:PhiU>PhiD}. We propose a linear model for $\lambda(\phi_+, \phi_-)$ as follows
\begin{equation}
  \lambda = a_0 + a_+\phi_+ + a_- |\phi_-| + \epsilon,
\end{equation}
where $\epsilon$ is the noise term. The regression model proves excellent since the adjusted R-square is 0.98. We obtain that $a_0=8.72\times10^{-8}$ with a $p$-value of 0.92, $a_+=9.91\times10^{-5}$ with a $p$-value of 0.0000, and $a_-=-9.19\times10^{-5}$ with a $p$-value of 0.0000. It indicates that the constant term $a_0$ is statistically equal to 0, and $\phi_+$ and $|\phi_-|$ have significant impacts on the divergence rate $\lambda$. The divergence rate $\lambda$ increases with $\phi_+$ and decreases with $|\phi_-|$. In addition, the impact is higher for $\phi_+$ since $a_+>|a_-|$.

\begin{table}[htp]
  \centering
  \caption{Estimated divergence rates of the price trajectories when $|\phi_-|<\phi_+$ and the half-lives $t_{1/2}$ when $|\phi_-|>\phi_+$. The first row shows the $\phi_+$ values, whereas the first column shows the $\phi_-$ values. The upper triangle presents the convergence rates $\lambda$ (multiplied by $10^5$), while the lower triangle presents the half-lives $t_{1/2}$ (multiplied by $10^{-4}$).}
  \medskip
  \label{Tb:EBOD:ComputExper:Price:PhiU>PhiD}
  \centering
  \begin{tabular}{ccccccc}
  \hline \hline
         & 0.05 & 0.10  & 0.15  & 0.20  & 0.25  & 0.30   \\
  \hline
    -0.05 &       & 0.520 & 1.050 & 1.630 & 2.041 & 2.508  \\
    -0.10 & 10.80 &       & 0.543 & 1.063 & 1.619 & 1.977  \\
    -0.15 & 4.467 & 6.885 &       & 0.593 & 1.117 & 1.648  \\
    -0.20 & 2.828 & 4.477 & 1.688 &       & 0.562 & 1.000  \\
    -0.25 & 2.588 & 4.971 & 5.662 & 9.713 &       & 0.854  \\
    -0.30 & 1.866 & 1.965 & 2.871 & 3.366 & 7.983 &   \\
  \hline\hline
 \end{tabular}
\end{table}

In order to quantify the decay rate of the price for the cases of $|\phi_-|>\phi_+$, we calculate the half-lives $t_{1/2}$. The half-life is the average of time moments that intersect the horizontal line $p_t=p_1/2=5$. The half-lives $t_{1/2}$ are digested in the lower triangle of Table \ref{Tb:EBOD:ComputExper:Price:PhiU>PhiD}. We also propose a linear model for $t_{1/2}(\phi_+, \phi_-)$ as follows
\begin{equation}
  t_{1/2} = b_0 + b_+\phi_+ + b_- |\phi_-| + \epsilon,
\end{equation}
The regression model proves good since the the adjusted R-square is 0.50. We obtain that $b_0=1.22\times10^{5}$ with a $p$-value of 0.001, $b_+=4.56\times10^{5}$ with a $p$-value of 0.007, and $b_-=-5.04\times10^{5}$ with a $p$-value of 0.004. It indicates that $\phi_+$ and $\phi_-$ have significant impacts on the half-life $t_{1/2}$. The half-life $t_{1/2}$ also increases with $\phi_+$ and decreases with $|\phi_-|$. In addition, the impact is higher for $\phi_-$ since $|b_-|>b_+$.

Our computations provide technical evidence against the asymmetric price limit rule (Article VII) proposed in the  {\textit{Draft Guideline on the Trading of Stocks Bearing Risk Warnings}}. The results show that, with $\phi_-=-0.05$ and $\phi_+=0.02$, the price of a stock labelled with risk warnings will gradually vanish. These risk-warning stocks will eventually delisted from the SHSE.

\subsection{Return distributions}

\begin{figure}[tb]
  \centering
  \includegraphics[width=8cm,height=5.5cm]{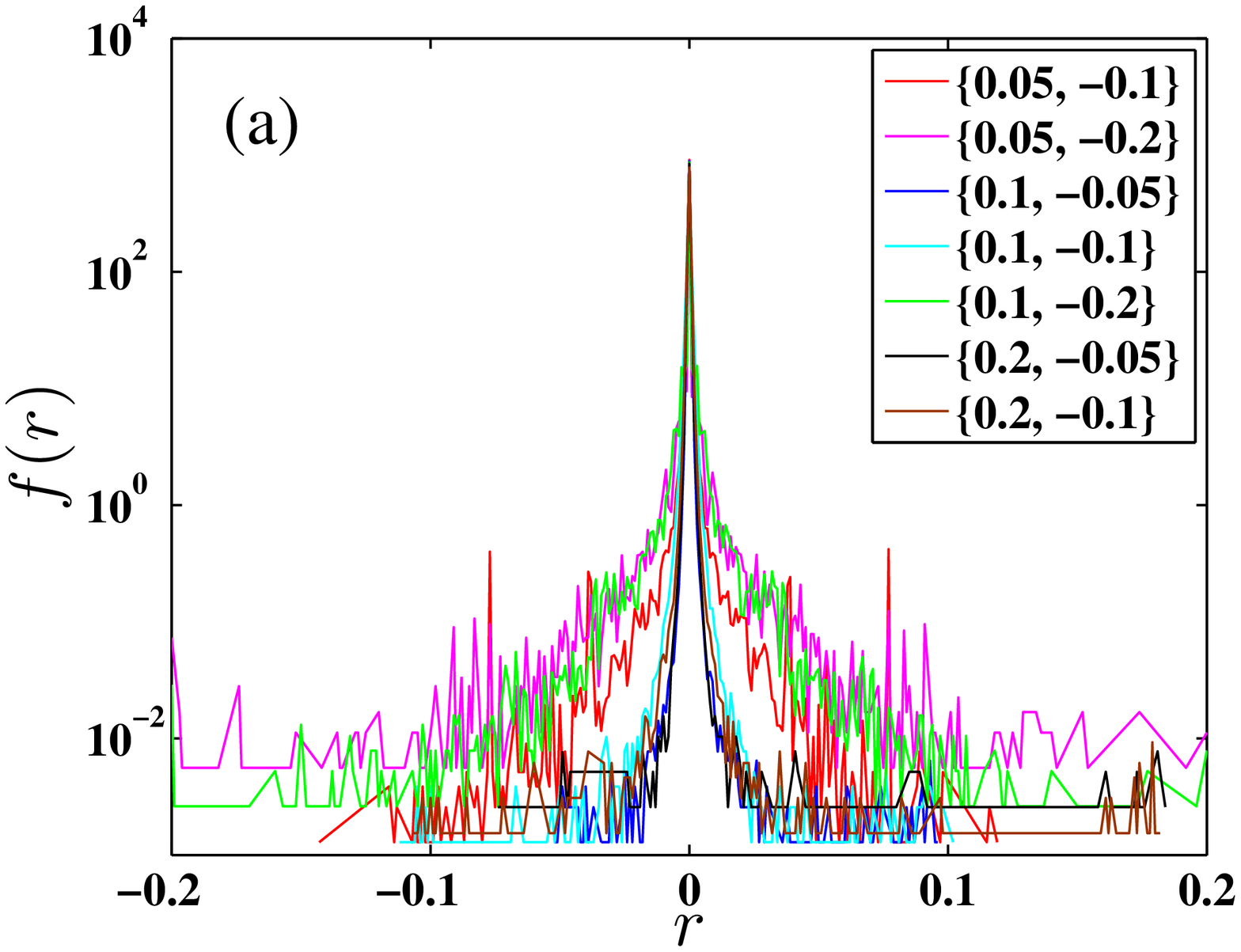}
  \includegraphics[width=8cm,height=5.5cm]{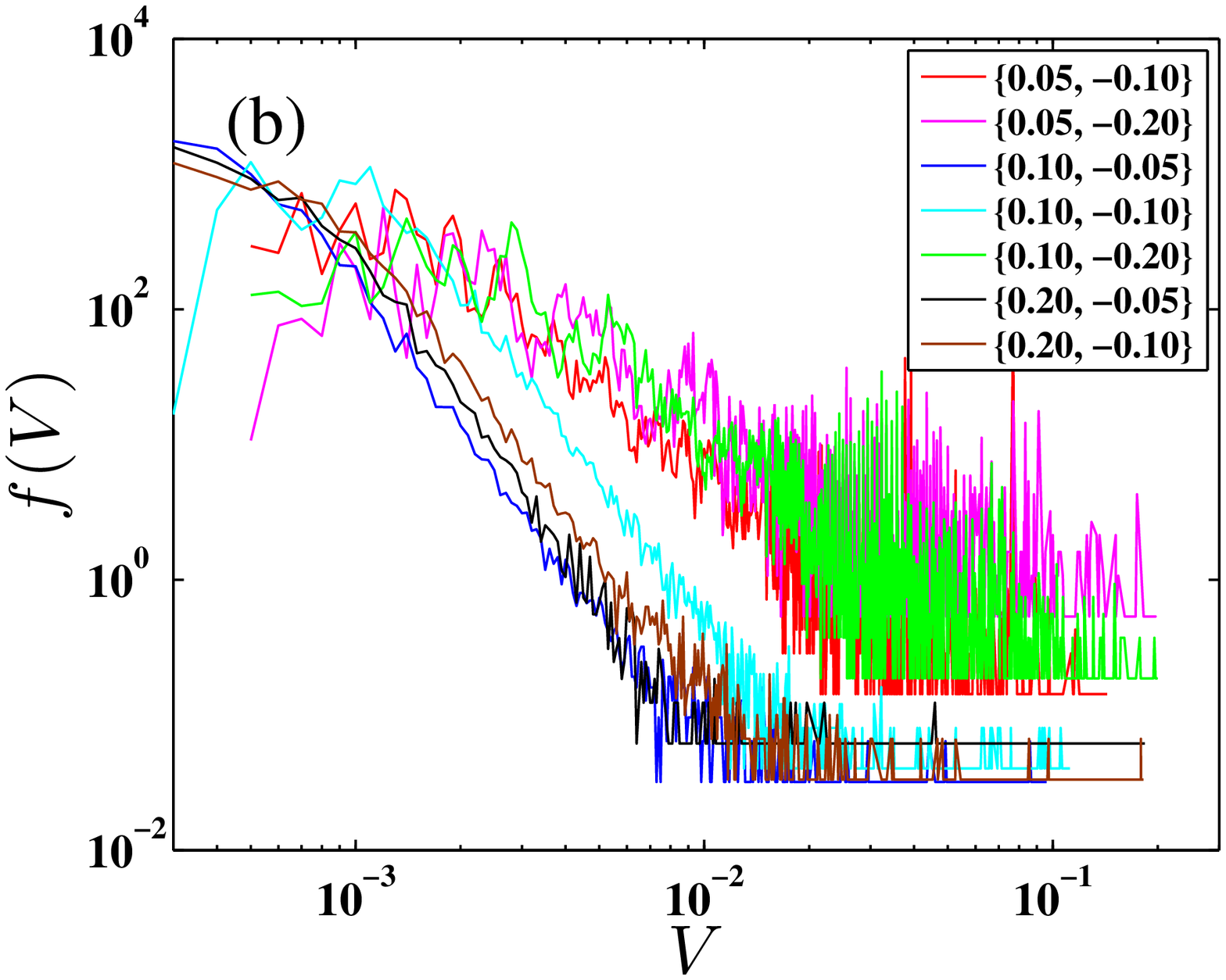}%
  \caption{(color online) Empirical distributions of returns in linear-log coordinates (a) and absolute returns in log-log scales (b) for the stocks simulated for different combinations of price limits $\{\phi_+,\phi_-\}$.}
  \label{Fig:ODM:Experiment:PDF}
\end{figure}

With several example combinations of price limits $\{\phi_+,\phi_-\}$, Fig.~\ref{Fig:ODM:Experiment:PDF}(a) shows the empirical distributions of returns, while Fig.~\ref{Fig:ODM:Experiment:PDF}(b) illustrates the empirical distributions of absolute returns, which have evident power-law tails. We unveil two clusters of distributions. The distributions for $|\phi_-|\le\phi_+$ (first group) are relatively narrow in the bulks and their tail exponents have comparable values. The distributions for $|\phi_-|>\phi_+$ (second group) are relatively broad in the bulks and their tail exponents are also comparable but different from the first group.

\begin{table}[htp]
  \centering
  \caption{Estimated tail exponents of the simulated returns for different price limit combinations. The first row shows the $\phi_+$ values, whereas the first column shows the $\phi_-$ values.}
  \medskip
  \label{Tb:EBOD:ComputExper:Return:alpha}
  \centering
  \begin{tabular}{cccccccccc}
  \hline \hline
         & 0.05 & 0.10  & 0.15  & 0.20  & 0.25  & 0.30   \\
  \hline
  -0.05 &  3.32 &  3.36 &  3.37 &  3.44 &  3.48 &  3.53 \\
  -0.10 &  1.94 &  3.41 &  3.51 &  3.43 &  3.52 &  3.50 \\
  -0.15 &  1.92 &  2.02 &  3.34 &  3.53 &  3.54 &  3.53 \\
  -0.20 &  1.96 &  1.99 &  2.02 &  3.41 &  3.54 &  3.54 \\
  -0.25 &  1.91 &  2.08 &  2.03 &  2.12 &  3.20 &  3.45 \\
  -0.30 &  2.01 &  2.02 &  2.11 &  2.06 &  2.11 &  2.72 \\
  \hline\hline
 \end{tabular}
\end{table}

The presence of power-law tails observed in Fig.~\ref{Fig:ODM:Experiment:PDF}(b) is confirmed by the method of \cite{Clauset-Shalizi-Newman-2009-SIAMR}. The estimated tail exponents for all the price limit combinations under investigation are depicted in Table \ref{Tb:EBOD:ComputExper:Return:alpha}. Consistent with Fig.~\ref{Fig:ODM:Experiment:PDF}, the tail exponents can be divided into two groups, in which $\langle\alpha\rangle=2.02\pm0.07$ for $|\phi_-|>\phi_+$ and $\langle\alpha\rangle=3.41\pm0.18$ for $|\phi_-|\le\phi_+$. To further understand the impacts of $\alpha(\phi_+,\phi_-)$, we regress the following linear equation:
\begin{equation}
  \alpha = c_0 + c_+\phi_+ + c_- |\phi_-| + \epsilon.
\end{equation}
For all the combinations of price limits $\{\phi_+,\phi_-\}$, we obtain that $c_0= 2.85$, $c_+= 4.88$, and $c_-=-4.98$, whose $p$-values are all less than 0.0000. The adjusted R-square is  0.70 and the MSE is 0.1503. Although all the coefficients are significantly different from 0 and the goodness-of-fit is high, it is equivalently fitting ``two points'' due to the clear separation of the tail exponents and the closeness of the tail exponents in each group.
For the combinations with $|\phi_-|>\phi_+$, we obtain that $c_0= 1.87$ with a $p$-value of 0.0000, $c_+= 0.75$ with a $p$-value of 0.0015, and $c_-= 0.27$ with a $p$-value of 0.1748. The adjusted R-square is  0.69 and the MSE is 0.0015.
For the combinations with $|\phi_-|\le\phi_+$, we obtain that $c_0= 3.34$ with a $p$-value of 0.0000, $c_+= 0.53$ with a $p$-value of 0.0429, and $c_-= 0.18$ with a $p$-value of 0.4491. The adjusted R-square is 0.36 and the MSE is 0.0024.
The results suggest that an increase of $|\phi_-|$ or $\phi_+$ will increase the tail exponent $\alpha$, expect for $\phi_-$ of the $|\phi_-|\le\phi_+$ case. This finding is reasonable since wider price limits will leads to more return points with larger magnitudes.

In Table \ref{Tb:EBOD:ComputExper:Return}, we present the average returns for different price limit combinations. When $|\phi_-|<\phi_+$, the average returns are positive. When $|\phi_-|>\phi_+$, the average returns are negative. When $|\phi_-|=\phi_+$, the average return decreases monotonically from positive to negative with the increase of $\phi_+$. We use the following linear model to fit the data
\begin{equation}
  \langle{r}\rangle = d_0 + d_+\phi_+ + d_- |\phi_-| + \epsilon
\end{equation}
and obtain that $d_0=0.0041\times10^{-4}$ with a $p$-value of 0.7363, $d_+=1.2380\times10^{-4}$ with a $p$-value of 0.0000, and $d_-=-1.4147\times10^{-4}$ with a $p$-value of 0.0000. The adjusted R-square is 0.98, indicating an excellent linear relationship between the average return and the two price limits.

\begin{table}[htp]
  \centering
  \caption{Average returns for different price limit combinations. The first row shows the $\phi_+$ values, whereas the first column shows the $\phi_-$ values. The values of the average returns have been multiplied by $10^5$ for better presentation.}
  \medskip
  \label{Tb:EBOD:ComputExper:Return}
  \centering
  \begin{tabular}{cccccccccc}
  \hline \hline
         & 0.05 & 0.10  & 0.15  & 0.20  & 0.25  & 0.30   \\
  \hline
    -0.05 & 0.017 & 0.544 & 1.098 & 1.721 & 2.188 & 2.550 \\
  -0.10 & -0.558 & 0.008 & 0.551 & 1.087 & 1.666 & 1.950 \\
  -0.15 & -1.388 & -0.746 & -0.022 & 0.582 & 1.144 & 1.674 \\
  -0.20 & -2.181 & -1.384 & -0.719 & -0.052 & 0.462 & 0.934 \\
  -0.25 & -2.990 & -2.158 & -1.433 & -0.843 & -0.190 & 0.494 \\
  -0.30 & -4.003 & -3.593 & -2.563 & -2.121 & -1.056 & -0.324 \\
  \hline\hline
 \end{tabular}
\end{table}

\subsection{Hurst exponents of returns}

We perform DMA analysis on the return time series for all the combinations of price limits. The estimated Hurst exponents are presented in Table \ref{Tb:EBOD:ComputExper:Return:H}. Some characteristic features can be derived from the table, which can be well captured by the linear model:
\begin{equation}
  H_r = e_0 + e_+\phi_+ + e_- |\phi_-| + \epsilon.
\end{equation}
For the combinations with $|\phi_-|>\phi_+$, we obtain that $e_0= 0.40$ with a $p$-value of 0.0000, $e_+= 0.01$ with a $p$-value of 0.3912, and $e_-= 0.30$ with a $p$-value of 0.0000. The adjusted R-square is 0.97 and the MSE is 0.000010. Hence, $H_r$ does not depend on $\phi_+$ since its coefficient is insignificant, but on $|\phi_-|$. The Hurst exponent of returns increases with $|\phi_-|$.
For the combinations with $|\phi_-|<\phi_+$, we obtain that $e_0= 0.49$ with a $p$-value of 0.0000, $e_+= 0.27$ with a $p$-value of 0.0000, and $e_-=-0.03$ with a $p$-value of 0.4598. The adjusted R-square is  0.80 and the MSE is 0.000066. It shows that $H_r$ does not depend on $\phi_-$. Rather, the Hurst exponent of returns increases with $|\phi_+|$.
For the combinations with $|\phi_-|=\phi_+$, we observe a humped shape for the dependence of $H_r$ with respect to $\phi_+$. Despite of the statistically significant trends, the Hurst indexes are all close to 0.5. The slight deviations are reasonable because of the non-arbitrage nature of the intraday returns caused by the $T+1$ trading mechanism \citep{Zhou-Gu-Jiang-Xiong-Chen-Zhang-Zhou-2017-CE}.

\begin{table}[htp]
  \centering
  \caption{Estimated Hurst exponents of the simulated returns for different price limit combinations. The first row shows the $\phi_+$ values, whereas the first column shows the $\phi_-$ values.}
  \medskip
  \label{Tb:EBOD:ComputExper:Return:H}
  \centering
  \begin{tabular}{cccccccccc}
  \hline \hline
         & 0.05 & 0.10  & 0.15  & 0.20  & 0.25  & 0.30   \\
  \hline
  -0.05 & $ 0.45\pm 0.01$ & $ 0.50\pm 0.01$ & $ 0.53\pm 0.01$ & $ 0.55\pm 0.02$ & $ 0.56\pm 0.01$ & $ 0.57\pm 0.01$ \\
  -0.10 & $ 0.43\pm 0.01$ & $ 0.50\pm 0.02$ & $ 0.53\pm 0.01$ & $ 0.55\pm 0.01$ & $ 0.56\pm 0.01$ & $ 0.56\pm 0.02$ \\
  -0.15 & $ 0.45\pm 0.01$ & $ 0.45\pm 0.01$ & $ 0.52\pm 0.02$ & $ 0.55\pm 0.01$ & $ 0.56\pm 0.02$ & $ 0.57\pm 0.01$ \\
  -0.20 & $ 0.46\pm 0.01$ & $ 0.46\pm 0.01$ & $ 0.47\pm 0.01$ & $ 0.54\pm 0.02$ & $ 0.55\pm 0.02$ & $ 0.56\pm 0.01$ \\
  -0.25 & $ 0.48\pm 0.02$ & $ 0.48\pm 0.01$ & $ 0.48\pm 0.01$ & $ 0.48\pm 0.01$ & $ 0.54\pm 0.02$ & $ 0.56\pm 0.02$ \\
  -0.30 & $ 0.49\pm 0.01$ & $ 0.50\pm 0.02$ & $ 0.49\pm 0.01$ & $ 0.50\pm 0.01$ & $ 0.49\pm 0.01$ & $ 0.52\pm 0.03$ \\
  \hline\hline
 \end{tabular}
\end{table}

\subsection{Hurst exponents of volatilities}

We also perform DMA analysis on the volatility time series for all the combinations of price limits. The estimated Hurst exponents are presented in Table \ref{Tb:EBOD:ComputExper:Volatility:H}. We also find some intriguing features, which can be well captured by the linear model:
\begin{equation}
  H_V = f_0 + f_+\phi_+ + f_- |\phi_-| + \epsilon.
\end{equation}
For the combinations with $|\phi_-|\ge\phi_+$, we obtain that $f_0= 0.68$ with a $p$-value of 0.0000, $f_+= 0.16$ with a $p$-value of 0.0001, and $f_-= 0.13$ with a $p$-value of 0.0006. The adjusted R-square is 0.80 and the MSE is 0.000088. In this case, the Hurst exponent of volatility increases with $\phi_+$ and $|\phi_-|$ and the impact of $\phi_+$ is larger than $|\phi_-|$.
For the combinations with $|\phi_-|<\phi_+$, we obtain that $f_0= 0.77$ with a $p$-value of 0.0000, $f_+=-0.02$ with a $p$-value of 0.6775, and $f_-=-0.01$ with a $p$-value of 0.8870. The adjusted R-square is -0.13 and the MSE is 0.000094. In this case, the Hurst exponent $H_V$ is independent of $\phi_+$ and $|\phi_-|$ and remains constant.

\begin{table}[htp]
  \centering
  \caption{Estimated Hurst exponents of the simulated volatilities for different price limit combinations. The first row shows the $\phi_+$ values, whereas the first column shows the $\phi_-$ values.}
  \medskip
  \label{Tb:EBOD:ComputExper:Volatility:H}
  \centering
  \begin{tabular}{cccccccccc}
  \hline \hline
         & 0.05 & 0.10  & 0.15  & 0.20  & 0.25  & 0.30   \\
  \hline
  -0.05 & $ 0.71\pm 0.01$ & $ 0.78\pm 0.01$ & $ 0.77\pm 0.01$ & $ 0.75\pm 0.02$ & $ 0.75\pm 0.02$ & $ 0.77\pm 0.01$ \\
  -0.10 & $ 0.68\pm 0.01$ & $ 0.72\pm 0.01$ & $ 0.76\pm 0.02$ & $ 0.76\pm 0.02$ & $ 0.76\pm 0.02$ & $ 0.76\pm 0.01$ \\
  -0.15 & $ 0.70\pm 0.01$ & $ 0.71\pm 0.02$ & $ 0.72\pm 0.02$ & $ 0.75\pm 0.01$ & $ 0.77\pm 0.03$ & $ 0.77\pm 0.01$ \\
  -0.20 & $ 0.70\pm 0.02$ & $ 0.71\pm 0.01$ & $ 0.73\pm 0.02$ & $ 0.75\pm 0.01$ & $ 0.75\pm 0.02$ & $ 0.76\pm 0.01$ \\
  -0.25 & $ 0.73\pm 0.02$ & $ 0.73\pm 0.02$ & $ 0.73\pm 0.02$ & $ 0.74\pm 0.02$ & $ 0.75\pm 0.02$ & $ 0.76\pm 0.02$ \\
  -0.30 & $ 0.73\pm 0.02$ & $ 0.75\pm 0.02$ & $ 0.73\pm 0.03$ & $ 0.75\pm 0.02$ & $ 0.75\pm 0.02$ & $ 0.77\pm 0.02$ \\
  \hline\hline
 \end{tabular}
\end{table}

\section{Conclusion}
\label{sec:conclusion}

We have developed a behavioural order-driven model with price limit rules based on the empirical regularities of the order placement and cancellation processes. In the order placement process, order directions are determined by the unveiled long memory, order prices are determined by the long memory in relative prices and the asymmetric distribution of relative prices, and order sizes are determined by the nonlinear dependence on the relative price. In the order cancellation process, we adopted a Poisson process with the arrival rate determined from real data and the cancelled order is determined empirically by its temporal and spatial positions. The model is validated because it can reproduce the main stylized facts in real markets.

Computational experiments uncover that asymmetric setting of price limits will cause the stock price diverging when the up limit $\phi_+$ is greater than the absolute down limit $|\phi_-|$ and vanishing vice versus. When $\phi_+>|\phi_-|$, the price diverges exponentially and the divergence rate increases linearly with $\phi_+$ and decreases linearly with $|\phi_-|$. When $\phi_+<|\phi_-|$, the price will eventually decay to a lower bound that is determined by $|\phi_-|$ and the tick size. According to these results, setting asymmetric price limits will destroy the market.

For asymmetric price limits, the simulated returns have power-law tails. The tail exponents for $\phi_+<|\phi_-|$ are significantly smaller than those for $\phi_+\ge|\phi_-|$. In addition, the tail exponent $\alpha$ increases linearly with $\phi_+$ and decreases linearly with $|\phi_-|$. Interestingly, the average return increases linearly with $\phi_+$ and decreases linearly with $|\phi_-|$. The Hurst exponents $H_r$ of returns are all close to 0.5, but have slight deviations. For each case of asymmetric price limits, the Hurst exponent depends linearly on the price limits. In both cases, $H_r$ increases with $\phi_+$ and $|\phi_-|$. For volatility, the Hurst exponents $H_V$ are constant for $\phi_+>|\phi_-|$, and $H_V$ increases with $\phi_+$ and $|\phi_-|$ for $\phi_+<|\phi_-|$.

Our EBOD model provides a suitable computational experiment platform for academics, market participants and policy makers \citep{Farmer-Foley-2009-Nature}. For academics, the model can be used to study microstructure theories such as the price impacts of transactions, influencing factors of macroscopic properties, the mechanisms of the formation of bubbles and the triggering of crashes through collective herding behaviours, systemic risks, and so on. For market participants, the model can be used to design optimal strategies of order placement, perform option pricing, estimate value-at-risk with higher precision, and predict reoccurrence probability of extreme fluctuations. For policy makers, the model can be use to study the effects of different setting of price limits on the performance of markets, such as price discovery, market volatility and market liquidity. Certainly, the model is still open for further improvements following further efforts of model calibration.

\section*{Acknowledgements}

This work was partly supported by National Natural Science Foundation of China (Grants No. 71671066, 71501072, 71571121 and 71532009), and the Fundamental Research Funds for the Central Universities (222201718006).


\bibliographystyle{elsarticle-harv}
\bibliography{E:/Papers/Auxiliary/Bibliography}

\end{document}